\documentclass[12pt,a4paper,jmp,nofootinbib]{revtex4-1}

\usepackage{graphicx}
\usepackage{amsmath}
\usepackage{amssymb,amsfonts}
\usepackage{bbold}
\usepackage{hyperref}
\usepackage{xargs}
\usepackage{mathtools}
\usepackage{tikz}
\usepackage{tikz-cd}
\usepackage{mathrsfs}
\usepackage{eucal}
\usepackage{physics}
\usepackage{amsthm}

\usetikzlibrary{arrows.meta} 
\usetikzlibrary{calc}

\usepackage[margin=2.2cm]{geometry}

\newcommand{\R}{\mathbb{R}}
\newcommand{\C}{\mathbb{C}}
\renewcommand{\P}{\mathbb{P}}
\newcommand{\Z}{\mathbb{Z}}

\newcommand{\cO}{\mathcal{O}}

\newcommand{\bZ}{\mathbb{Z}}

\def\nfoot#1{{\begingroup\linespread{0.75}\footnote{#1}\endgroup}}

\newcommand{\thetabar}{\overline{\theta}}

\newcommand{\Q}{{Q}}
\newcommand{\Qbar}{\overline{{Q}}}
\newcommand{\Dbar}{\overline{\mathscr{D}}}
\newcommand{\N}{\mathcal{N}}
\newcommand{\alphadot}{\dot{\alpha}}
\newcommand{\betadot}{\dot{\beta}}

\usepackage{xparse}
\NewDocumentCommand \deq { o m }{
\begin{equation}
#2
\IfNoValueF{#1}{\label{#1}}
\end{equation}
}

\def\so{\ensuremath{\mathfrak{so}}}
\def\su{\ensuremath{\mathfrak{su}}}
\def\lie#1{\mathfrak{#1}}
\DeclareMathOperator{\Spin}{Spin}

\def\epar{\ensuremath{\varepsilon}}

\DeclareMathOperator{\Cl}{Cl}

\DeclareMathOperator{\Ext}{Ext}

\DeclareMathOperator{\ch}{ch}

\DeclareMathOperator{\Sym}{Sym}
\DeclareMathOperator{\graph}{graph}
\def\Nilpp(#1,#2){Y\qty({#1},{#2})}
\def\Nilp(#1,#2){\widehat{Y}\qty({#1},{#2})}
\def\Fun{\mathscr{O}}
\def\dual{\vee}

\def\dR#1#2{\frac{%
\partial^{\mathrlap{r}} \, {#1}}{\partial {#2}}}
\def\dL#1#2{\frac{%
\partial^{\mathrlap{l}} \, {#1}}{\partial {#2}}}
\def\af{+}
\DeclareMathOperator{\ad}{ad}
\def\ms{(-)}

\def\rep#1{\mathbf{#1}}
\def\crep#1{\smash{\mathbf{\overline{#1}}}}

\def\BVS{\mathfrak{S}}
\def\bvd{\mathfrak{s}}

\usepackage{stmaryrd}
\def\antibr#1{\left\llbracket{#1}\right\rrbracket}

\newtheorem{satz}{Fact}
\numberwithin{satz}{section}

\newlength{\myline}
\newcommandx*{\triplearrow}[4][1=0, 2=1]{
  \draw[line width=\myline,double distance=3\myline,#3] #4;
  \draw[line width=\myline,shorten <=#1\myline,shorten >=#2\myline,#3] #4;
}
\newcommandx*{\doublearrow}[4][1=0, 2=1]{
  \draw[line width=\myline,double distance=2\myline,#3] #4;
}
\newcommandx*{\quadarrow}[4][1=0, 2=2.5]{
  \draw[line width=\myline,double distance=5\myline,#3] #4;
  \draw[line width=\myline,double distance=\myline,shorten <=#1\myline,shorten >=#2\myline,#3] #4;
}

\tikzset{ 
my loop/.style={->,to path={ 
.. controls +(-45:2) and +(45:2) .. (\tikztotarget) \tikztonodes}}
} 

\tikzset{ 
left loop/.style={->,to path={ 
.. controls +(225:2) and +(135:2) .. (\tikztotarget) \tikztonodes}}
} 
\tikzset{ 
leftdown loop/.style={->,to path={ 
.. controls +(135:2) and +(225:2) .. (\tikztotarget) \tikztonodes}}
} 
\tikzset{ 
right loop/.style={->,to path={ 
.. controls +(-45:2) and +(45:2) .. (\tikztotarget) \tikztonodes}}
} 
\tikzset{ 
above loop/.style={->,to path={ 
.. controls +(45:2) and +(135:2) .. (\tikztotarget) \tikztonodes}}
} 
\tikzset{ 
below loop/.style={->,to path={ 
.. controls +(-135:2) and +(-45:2) .. (\tikztotarget) \tikztonodes}}
} 
\tikzset{
triple line/.code={\tikzset{%
        double equal sign distance, 
        double=\pgfkeysvalueof{/tikz/commutative diagrams/background color}}}
}

\numberwithin{subsubsection}{subsection}
\numberwithin{subsection}{section}
\numberwithin{equation}{section}

\makeatletter
\def\p@subsection{}
\def\p@subsubsection{}
\makeatother

\newcommand{\nocontentsline}[3]{}
\newcommand{\tocless}[2]{\bgroup\let\addcontentsline=\nocontentsline#1{#2}\egroup}

\begin{document}

\title{Holomorphic field theories and Calabi--Yau algebras}

\author{Richard Eager}
\email{eager@mathi.uni-heidelberg.de}
\author{Ingmar Saberi}
\email{saberi@mathi.uni-heidelberg.de}
\affiliation{Mathematisches Institut der Ruprecht-Karls-Universit\"at Heidelberg \\ Im Neuenheimer Feld 205 \\ 69120 Heidelberg, Germany}

\begin{abstract}
We consider the holomorphic twist of the worldvolume theory of flat D$(2k-1)$-branes transversely probing a Calabi--Yau manifold. 
A chain complex, constructed using the BV formalism, computes the local observables in the holomorphically twisted
theory.  Generalizing earlier work in the case $k=2$, we find that this complex can be identified with the Ginzburg dg algebra associated to the Calabi--Yau.
However, the identification is subtle; the complex is the space of fields contributing to the holomorphic twist of the free theory, and its differential arises from interactions. For $k=1$, this holomorphically twisted theory is related to the elliptic genus. We give a general description for D1-branes probing a Calabi--Yau fourfold singularity, and for $\mathcal{N}=(0,2)$ quiver gauge theories. 
In addition, we propose a relation between the equivariant Hirzebruch $\chi_y$ genus of large-$N$ symmetric products and cyclic homology.
\end{abstract}

\maketitle

\addtocounter{page}{-1}

\setcounter{tocdepth}{2}
\tableofcontents

\tocless\bibsection

\section{Introduction}

Supersymmetric gauge theories are a broad and interesting class of physical models. In addition to proposals that they may play a role in  physics beyond the Standard Model, studying them has led to much progress in our understanding of quantum field theories more generally over the last half century. 

One reason that supersymmetry has proved to be such a help in this arena is that supersymmetric theories often admit interesting truncations to particular sectors of states or operators, usually termed BPS~\cite{bogomolny,prasad}. Computations in these special sectors are usually much easier, as well as more robust under perturbations to the theory, than for the full interacting QFT. Various indices have been constructed that count BPS quantities, starting with the Witten index~\cite{WittenIndex}, which counts (with sign) the number of vacua in supersymmetric quantum mechanics, and is essentially analogous to the Euler characteristic of a chain complex. Many further indices have appeared in the literature~\cite[for example]{Witten:1986bf, CFIV, Romelsberger:2005eg, Kinney:2005ej}, and have been used heavily in the preceding years in the study of supersymmetric theories. 

Making a consistent truncation of a quantum field theory is subtle, and there is (as far as we are aware) essentially only one way to do it: one can take invariants of a symmetry that acts on the theory. The novelty of the BPS condition is that it requires states or operators to be invariant under a symmetry generator which is fermionic, rather than bosonic. In this case, the only appropriate notion of ``invariants'' appears when the generator is nilpotent, and produces the cohomology of that operator.

Another field that has seen great activity in recent years is that of \emph{topological quantum field theory}. Here, the idea is to construct consistent models of field theory which do not depend on the metric, and so produce topological (or smooth) invariants of the spacetime on which they are formulated. These models have been a source of fruitful interaction between theoretical physics, geometry, and topology. 

Topological field theories are customarily divided into two types: the so-called Schwarz type~\cite{WittenCS,SchwarzTQFT}, in which the action is explicitly independent of the metric, or the Witten type~\cite{WittenDonaldson}, which is a truncation (in exactly the fashion described above) of a full-fledged supersymmetric field theory. The subtlety here is that the choice of a nilpotent supercharge breaks the Lorentz symmetry; another action of the $\lie{so}(d)$ holonomy algebra on the fields must be found in order to formulate the field theory on a generic spacetime. This gives rise to the established name of the procedure, a ``topological twist.'' For an excellent review of topics in topological field theory, see~\cite{Birmingham:1991ty}.

As we have described it, the two procedures above seem to be almost the same, and the reader will wonder if in fact this is so. Roughly speaking, the answer is yes, up to the subtlety of which $Q$ is chosen: $\N=1$ theories in four dimensions, for example, have a well-defined superconformal index, but do not admit topological twists. In fact, the truncation of such a theory by~$Q$ is a peculiar sort of theory, intermediate between a standard and a topological QFT, that has been in the literature for some time~\cite{Witten:1991zz, Witten:1993yc, Park:1993ga, Johansen:1994aw, Witten:1994ev, Baulieu:2004pv, NekrasovThesis}, but has attracted new attention of late~\cite{Costello:2011np, Costello:2013zra, Costello-Scheimbauer,Costello:2016mgj,GGW}. 
These are the so-called \emph{holomorphic field theories;} the fields are holomorphic with respect to a chosen complex structure on the spacetime. One way to see this is to look at the $\N=1$ super-Poincar\'e algebra, and note that exactly half of the momenta (corresponding to antiholomorphic components) are $Q$-exact. Another is to note that,
in even dimensions, the choice of~$Q$ preserves a subalgebra $\lie{u}(n)\subseteq \lie{so}(2n)$ of the Lorentz algebra, with respect to which it is a (charged) scalar. The theory can thus be formulated on 
Calabi--Yau manifolds. 
Since such manifolds admit Killing spinors, the supersymmetry $Q$ is not broken by the non-flat spacetime, even in the absence of twisting (as would generically be the case). 

An important class of supersymmetric field theories arise from D-branes wrapping supersymmetric (BPS) cycles~\cite{Becker:1995kb, Bershadsky:1995qy, Ooguri:1996ck}. The two types of supersymmetry-preserving cycles in Calabi--Yau varieties are, loosely speaking, special Lagrangians and holomorphic cycles.  Since we are interested in holomorphic field theories, it is natural to focus on holomorphic cycles. More precisely, the worldsheet description of B-branes as boundary conditions in topological open string theory on Calabi--Yau varieties leads to the identification of the BPS boundary conditions as objects in the derived category of coherent sheaves on the Calabi--Yau variety~\cite{Sharpe:1999qz, Douglas:2000gi}.

In the context of geometric engineering and the AdS/CFT duality, D-branes wrapping cycles in non-compact Calabi--Yau varieties play a prominent role.  The open strings have a simple description in terms of quiver gauge theories \cite{Douglas:1996sw}.  The category of branes in this case has a simple description in terms of quivers with superpotential.  The superpotential arises from the identification of the topological open string theory with a holomorphic analog of Chern--Simons theory \cite{Witten:1992fb, Aspinwall-Katz}.  Berenstein and Douglas~\cite{Berenstein:2002fi} realized that the Calabi--Yau condition translates into a noncommutative analog of Serre duality on the quiver.  This notion was formalized as a Calabi--Yau $d$-algebra by Ginzburg~\cite{Ginzburg:2006fu}.  
Additional motivation for studying these algebras was related to mirror symmetry---roughly, a Calabi--Yau algebra has a derived category of finitely-generated projective modules that ``has all the essential features'' of the derived category of coherent sheaves on a Calabi--Yau variety~\cite[\S3]{Ginzburg:2006fu}.

One goal of this work is to describe the holomorphic twists of the supersymmetric gauge theories arising from D-branes wrapping BPS cycles and in particular the worldvolume theory of D-branes probing the tip of a Calabi-Yau cone and their holomorphic observables.  When the worldvolume theory flows to a superconformal field theory at low energies a particularly interesting observable is the superconformal index.  We will see that the superconformal index is a holomorphic observable.  The claim to fame of the superconformal index is that it is robust invariant of strongly coupled field theories that can often be computed using free field methods \cite{Romelsberger:2005eg, Kinney:2005ej}.  The large-$N$ four dimensional superconformal index of the worldvolume theory D-branes probing a Calabi-Yau singularity was shown to equal that of its holographic dual in \cite{Eager:2012hx}, building on earlier work \cite{Kinney:2005ej, Nakayama:2006ur, Gadde:2010en}.  Relatedly, a holomorphically twisted form of the AdS/CFT correspondence was proposed by Costello and Li~\cite{Costello:2016mgj}.  We will explain how to recover supersymmetric indices from the twisted AdS/CFT correspondence.

In this work we will consider the spectrum of holomorphic observables both for finite-rank gauge groups and in the large-$N$ limit.  As in~\cite{Kinney:2005ej,Eager:2012hx}, there will be considerable simplification at large $N$.  
The holomorphically twisted theory can, in some sense, be seen as (at least a first step toward) a categorification of the superconformal index.  See~\cite{Grant:2008sk, Chang:2013fba} for important earlier work in this direction.

A key ingredient of the matching of the superconformal index under the AdS/CFT duality was the ability to mimic the passage from open to closed strings by relating the superconformal index to the cyclic homology of a
differential graded algebra corresponding to the Calabi--Yau singularity.
 Ginzburg associated this dg algebra to every Calabi--Yau $d$-algebra; we refer to it as the Ginzburg dg algebra.  For Calabi--Yau 3-algebras, it was shown in~\cite{Eager:2012hx} that the cyclic homology of the Ginzburg dga (associated, for example, to a quiver with superpotential) corresponds to a space of states in four-dimensional (quiver) gauge theory. Its differential corresponds to the supercharge that is used to define the superconformal index.
Here, the general setting was that of compactifications of type IIB string theory of the form $\R^4 \times X$, with $X$ a Calabi--Yau threefold.  It is thus natural to expect that for Calabi--Yau 4 algebras, the cyclic homology of the Ginzburg dga can be used to compute the large-$N$ elliptic genus.  We will see much evidence for this expectation; however, a few new subtleties arise.

The method of computing the index used in~\cite{Eager:2012hx} first imposed a BPS bound and identified letters that could contribute to the index, and then took cohomology of a further ``supercharge'' to produce the index itself (in the process imposing the condition of gauge invariance). It was the intermediate space, generated by particular letters, that was identified with Ginzburg's dg algebra.

One key point of the current work is that this procedure, which may seem somewhat \emph{ad~hoc}, in fact has a precise physical meaning. Namely, one first computes the space of BPS operators in the \emph{free} theory, and then perturbs away from this to the BPS operators in the interacting theory itself. Indeed, in~\cite{Gukov:2015gmm}, it was observed that one can formulate a twist of a Lagrangian supersymmetric theory in terms of a bicomplex in several ways: one can either think of the twisted theory as arising from truncating the free theory (corresponding to the decomposition $d = \bvd + Q$ of the differential into the original BRST part and a part from the supersymmetry algebra), or one can imagine perturbing from the twisted free theory to the twisted interacting theory, and correspondingly decompose $d$ as $d_0 + d_1$.
Thus, one should generally expect a spectral sequence from the BPS states or operators of a theory at zero coupling to those of the theory at finite coupling, and this spectral sequence was shown to correspond to the cancellations occurring in the usual elliptic genus computations for the Landau--Ginzburg model in two dimensions \cite{Witten:1993jg}. A similar spectral sequence is implicitly at work in the computations in~\cite{Chang:2013fba,Eager:2012hx}.

Recognizing this allows us to repeat the same analysis for the worldvolume theory of D$1$-branes transverse to a Calabi--Yau fourfold. We identify the Ginzburg algebra corresponding to the Calabi--Yau fourfold with the complex consisting of the cohomology of~$d_0$, equipped with the differential $d_1$. (Implicit in this is the assumption that the spectral sequence collapses at the $E_2$ page.) The relevant index is the elliptic genus of the two-dimensional theory; for $\N=(8,8)$ Yang--Mills, this was computed in detail in~\cite{Murat}. 
While we only discuss the example of maximally supersymmetric Yang--Mills theory in detail, it should be straightforward to find analogous results for any theory of gauge fields with adjoint matter, and (we suspect) for quiver $(0,2)$ theories in general. Unlike for threefolds, it is important to recognize which spectral sequence is in play: the differential in the Ginzburg algebra, $d_1$, contains terms originating both in~$Q$ and in~$\bvd$, and cannot be matched otherwise.

The worldvolume theory on D1-branes at the tip of a Calabi--Yau cone generically only preserves $\N=(0,2)$ supersymmetry.  The worldvolume theory had been described for quotient singularities in \cite{Mohri:1997ef, GarciaCompean:1998kh}.  While several mathematical works \cite{Craw:2010zi, Lam:2014} describe generic CY 4-algebras arising from Calabi--Yau cones, the first explicit connection between CY 4-algebras and $(0,2)$ theories was made in \cite{EagerCY4}.  While two-dimensional $\N=(0,2)$ theories are typically described using $E$ and $J$-type potentials with $E_aJ^a = 0$, as reviewed in section \ref{sec:0,2}, CY 4-algebras are often described with a potential in degree $-1.$  The key point is that the degree-$(-1)$ potential encodes both the $E$ and $J$-terms.  However, the cohomology problem we conisder can be adapted to arbitrary $(0,2)$ gauge theories, such as those arising from M5-branes wrapping four-manifolds \cite{Gadde:2013sca} or F-theory compactifications \cite{Schafer-Nameki:2016cfr}.

Perhaps the earliest appearance of holomorphically twisted theories are the ``half-twisted'' theories in two dimensions \cite{Witten:1991zz, Witten:1993yc}.  The partition function for the half-twisted
Landau-Ginzburg model is its elliptic genus.  It was computed in \cite{Witten:1993jg} using a free field realization that is a variant of a free first-order $(\beta, \gamma)$ system \cite{Fre:1992hp}. 
The four dimensional ``big brother'' of the elliptic genus is the superconformal index.  The four-dimensional superconformal index was computed for gauge theories in \cite{Romelsberger:2005eg, Kinney:2005ej} using a free field description colloquially known as letter counting.  Recently, the elliptic genus of two-dimensional gauge theories was computed in~\cite{Gadde:2013ftv} using letter counting and in~\cite{Benini:2013xpa,Benini:2013nda} using supersymmetric localization.

In four dimensions, the holomorphic twist of a chiral multiplet can also be described as a four-dimensional holomorphic $(\beta,\gamma)$ system~\cite{NekrasovThesis, Aganagic:2017tvx}.
Indeed, one of the things we would like to emphasize is that the four-dimensional superconformal index can be viewed as the partition function of a holomorphic $BF$ theory~\cite{Baulieu:2004pv, Costello:2013zra} coupled to a $(\beta, \gamma)$ system.  The somewhat \emph{ad~hoc} nature of determining which letters contribute to the index and which equations of motion have components on these letters is completely captured by the twisted theory.
 In this paper we provide the classical differential needed to compute the spectrum of states in the corresponding situation in two dimensions.  We plan to address quantum corrections to the differential in subsequent work. In the large-$N$ limit, we are able to relate this cohomology to cyclic homology, in the manner of~\cite{Eager:2012hx}.

For the sake of both self-containment and pedagogy, we rederive or rediscover a great many well-known results along the way, and provide reasonably detailed exposition thereof. No pretense of originality is made; the holomorphic twist of ten-dimensional Yang--Mills was first computed by Baulieu~\cite{Baulieu:2010ch}, who also provided the first analysis of the Batalin--Vilkovisky procedure for supersymmetric systems, showing how supersymmetry may be taken to act on antifields~\cite{Baulieu:1990uv}. The same twist was also considered in great detail by Costello and Li~\cite{Costello:2016mgj}, who generalized to maximally supersymmetric Yang--Mills theories in any even dimension. 
For an introduction to recent work on holomorphic twists see~\cite{Costello-Scheimbauer}.  See also~\cite{GGW} for some recent work.

Here is an outline of the rest of the paper. We summarize conventions in~\S\ref{sec:notation}, and review a bit of necessary representation theory in~\S\ref{sec:spinors}. \S\ref{sec:BV} is an expos\'e of necessary aspects of the Batalin--Vilkovisky formalism, and \S\ref{sec:twists} contains some general remarks on the idea of twisting. After these lengthy preliminaries, we study ten-dimensional super Yang--Mills theory in~\S\ref{sec:MSYM}, and perform its holomorphic twist in~\S\ref{sec:holo}. \S\ref{sec:dimred} considers dimensional reduction to maximally supersymmetric Yang--Mills in four and two dimensions.  \S\ref{sec:0,2} offers some quick review of $(0,2)$ theories, and \S\ref{sec:quivCY} discusses quivers and Calabi--Yau $d$-algebras. \S\ref{sec:DMVV} provides some comments on the large-$N$ limit of the proposed correspondence and connections to cyclic homology.

Note: After the results in \S\ref{sec:CYd} were announced in \cite{EagerCY4}, \cite{Closset:2017yte} appeared which provides further examples of 2d quivers arising from D$1$-branes at $\text{CY}^4$ singularities.

\subsection{Acknowledgments}

We thank K.~Costello, M.~Dedushenko, O.~Gwilliam, N.~Paquette, D.~Pei, R.~Plesser, A.~Tripathy, B.~Williams, and especially J.~Walcher for many useful conversations. R.E. would like to thank Perimeter Institute for hospitality during the earlier stages of this project.  R.E. thanks the organizers of the conferences ``Supersymmetric Theories, Dualities and Deformations'' at the Albert Einstein Center at the University of Bern and ``Mirror Symmetry and Applications'' at the Steklov Mathematical Institute of Russian Academy of Sciences and the	National Research University Higher School of Economics in Moscow, where preliminary versions of this work were first presented. I.A.S. is grateful to the Aspen Center for Physics and the Max Planck Institut f{\"u}r Mathematik for hospitality.  This research was supported in part by Perimeter Institute for Theoretical Physics. Research at Perimeter Institute is supported by the Government of Canada through Industry Canada and by the Province of Ontario through the Ministry of Economic Development and Innovation.  Our work is supported in part by the Deutsche Forschungsgemeinschaft, within the framework of the Exzellenzinitiative an der Universit{\"a}t Heidelberg. 

\section{Summary of conventions and notation} 
\label{sec:notation}

Throughout this paper, we will always work in Euclidean signature.
We will often use index notation, although indices may be suppressed (particularly from the spinor representation) when they can be restored unambiguously from the context. Indices will be written raised and lowered to indicate inequivalent complex conjugate representations; this convention will be disregarded for any representation whose indices can be meaningfully raised and lowered, such as the vector of $SO(2n)$.

A raised index for $SU(n)$ will denote the fundamental, and a lowered the antifundamental. Indices for $SU(5)$ will be taken from the set $jklmn$. There are two standard invariant tensors for $SU(5)$, which will be written $\delta^j_k$ and~$\epsilon_{jklmn}$.

$S_\pm$ will denote the spinor representation of $SO(2n)$, of positive or negative chirality; $D=S_+\oplus S_-$ will denote the Dirac spinor.  Note that the basis elements of $D$ will be numbered from $0$ to $2^n - 1$. A raised spinor index will denote $S_+$, and lowered $S_-$, of~$SO(10)$; such indices will be taken from the beginning of the Greek alphabet ($\alpha, \beta$, and so on). The gamma matrices and antisymmetrized products thereof are invariant tensors; $\Gamma$ will denote matrices acting in~$D$, which may have even or odd chirality, and $\gamma$ will denote the corresponding objects with chiral indices. Thus the standard gamma matrices of~$SO(10)$ are a pair of symmetric matrices, $\Gamma_\mu \sim \{ (\gamma_{\mu})_{ \alpha \beta}, \gamma_\mu^{\alpha \beta} \}$. We will write $\gamma$ only for emphasis. For more details on spinors, see the review in~\S\ref{sec:spinors} or~\cite{Deligne}.

Vector indices for $SO(2n)$ will come from the set $\mu,\nu,\ldots$. We will not place them with any particular care, although we do adhere to the summation convention for pairs of compatibly placed indices. Vector indices take values in $\{0,\ldots,2n-1\}$.

When necessary, we will try to further adhere to the convention that indices $a,b,\ldots$ refer to a basis for the adjoint representation of the gauge group, and indices $p,q,\ldots$ refer to $R$-symmetries appearing upon dimensional reduction. 

When discussing graded algebras (for instance, of polynomial expressions in the fields of a supersymmetric theory), it will be convenient to use left and right derivatives. The left derivative is the ordinary derivative, where the operator is imagined to act from the left; the right derivative differs from it by a sign, corresponding to the Koszul rule for switching the positions of the derivative and the expression. That is,
\deq{
\dR{y}{x} = \ms^{\abs{x} \cdot \abs{y}} \dL{y}{ x}.
}
(The absolute value signs here denote the parity of the corresponding expression.)

The \emph{nilpotent locus} of the $\N$-extended supersymmetry algebra in $d$ dimensions will be denoted $\Nilp(d,\N)$; it is a subspace of the complex vector space dual to the space of supersymmetry generators $Q$, where the Chevalley--Eilenberg ``ghosts'' take values. For instance, $\Nilp(10,1)\subset S_- \cong \C^{16}$. Since nilpotence is a scale-invariant condition, $\Nilp(d,\N)$ is always a complex cone, and descends to the projectivization of its ambient vector space; we denote its $\C^\times$ quotient by~$\Nilpp(d,\N)$.

 A very large number of nilpotent operators will appear in what follows, and it is both challenging and conceptually crucial to keep them all straight. In an effort to facilitate this task for the reader, we have endeavored to use a set of conventions that at least border on reason and consistency. In particular, 
departing somewhat from established standards, the letter $Q$ will only ever be used for a fermionic symmetry generator that originated in a physical supersymmetry algebra; it represents (whether before or after a twist) an element $u^\alpha Q_\alpha$ of the odd part of the super-Poincar\'e algebra, often with $u$ lying in $\Nilp(d,\N)$.

Nilpotent operators also appear in the context of gauge theories, completely independent of supersymmetry; for definitions and discussion of the concepts that appear, look ahead to~\S\ref{sec:BV}.
The BRST operator,  or Chevalley--Eilenberg differential, in a Koszul complex will be denoted $\kappa$. Note, though, that we reserve this for the differential encoding a gauge symmetry. We will similarly encode variations under supersymmetry, or other global algebras, in terms of a differential, which will in that case be denoted  $\delta$.

 In the BV formalism, two more nilpotent operators appear: the so-called ``BV Laplacian,'' which (in keeping with standard practice) we will call $\Delta$, and also the ``BV differential'' on operators, defined by the operation of antibracketing with the action:
\deq{
\bvd(\phi) = \antibr{\BVS, \phi}.
}
In the context of gauge theory, $s$ will be related to~$\kappa$, but is not necessarily the same in general! We remark also that, in the context of the BV formalism, we reserve $\antibr{\cdot,\cdot}$ for the antibracket, and~$\BVS$ for the BV action solving the quantum master equation. The superscript $(\cdot)^\af$ will label antifields.

$d$ will be used to represent any of these differentials, independent of its origin, which is being regarded at that time as part of the structure of a cochain complex. For example, after a twist, the relevant differential acting on the operators of a gauge theory will be of the form
\deq{d = \bvd + Q.}
But other decompositions of~$d$ may be useful. For example, we will have cause to separate $d$ into those terms appearing in the free theory (whether originating in $\bvd$ or~$Q$), and the terms whose presence is due to interactions. In this instance, we may write $d= d_0 + d_1 $. 

Lastly, in the pure spinor formalism for supersymmetric Yang--Mills theory, an entirely new differential also appears: this is Berkovits' operator $\mathscr{D}=\sum u^\alpha \mathscr{D}_\alpha$. $\mathscr{D}$ acts, not on the space of usual physical fields of Yang--Mills, but on a much larger space; it produces a BV complex that is equivalent to the (untwisted) usual one.

\section{Representation theory}
\label{sec:spinors}

\subsection{The algebra $\lie{so}(2n)$}

We will choose a basis for the Lie algebra $\so(2n)$ consisting of elementary antisymmetric matrices $M_{\mu\nu}$ (with $+1$ as the $\mu\nu$-th entry, $\mu<\nu$). It's then easy to check that 
\deq{
[M_{\mu\nu},M_{\lambda\rho}] = \delta_{\mu\lambda} M_{\nu\rho} + \delta_{\nu\rho} M_{\mu\lambda} - \delta_{\mu\rho}M_{\nu\lambda} - \delta_{\nu\lambda} M_{\mu\rho}.
}
We will choose notation that makes it convenient to think of the identification $\R^{2n} \cong \C^n$, in which the $j$-th complex factor corresponds to the $(2j,2j+1)$-plane in~$\R^{2n}$. In particular, we will choose the $k$-th Cartan generator to correspond to elementary rotations in the $(2k,2k+1)$-plane; that is,
\deq{
H_k = M_{2k,2k+1}, \qquad 0 \leq k \leq n-1.
}
The $H_k$ are therefore also a set of Cartan generators for the subgroup $U(n) \subseteq SO(2n)$ preserving our chosen complex structure.

We choose a basis of positive roots as follows:
\deq{
\alpha_k = e_k - e_{k+1} \text{ (for $0\leq k < n-1$);} \quad \alpha_{n-1} = e_{n-2} + e_{n-1}.
}
Here $e_k$ are the standard basis of~$\R^n$.
The fundamental weights corresponding to the two chiral spinor representations $S_\pm$ are:
\deq{
\mu_\pm = \frac{1}{2} \left( e_0 + \cdots + e_{n-2} \pm e_{n-1} \right).
}
When $n$ is odd, the representations are complex (and conjugate to one another), while for $n$ even they are either real or pseudo-real, depending on the value of $n\bmod 4$. In particular, the spinor representation in ten dimensions is complex.

\begin{figure}
\begin{center}
\begin{tikzpicture}
\tikzstyle{dynk}=[draw,scale=0.7,fill=white,minimum size=1.25 cm,circle]
\draw (0,0) node[dynk]{$\alpha_1$} -- (2,0) node (A) [dynk]{$\alpha_2$};
\draw (6,0) node (B) [dynk]{$\alpha_{n-2}$} -- ++(45:2) node[dynk]{$\alpha_{n-1}$};
\draw (B) -- ++(-45:2) node[dynk]{$\alpha_{n}$};
\draw[dashed] (A)  -- (B);
\end{tikzpicture}
\end{center}
\caption{Dynkin diagram and positive simple roots for~$\so(2n)$}
\label{fig:dynk}
\end{figure}
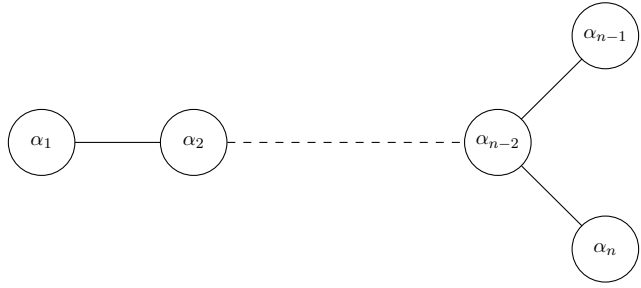

\subsection{Gamma matrices}

As is well-known, the Lie algebras of spin groups are related closely to Clifford algebras. We let the Euclidean Clifford algebra $\Cl_{2n}$ be generated by $\Gamma_\mu$ subject to the relation
\deq{
\{\Gamma_\mu, \Gamma_\nu \} = 2 \delta_{\mu\nu},
}
where the indices run from $0$ up to $2n-1$ as noted above.
 (There is an important reason for this convention; see below.) 
It is easy to see that $\Cl_{2n}$ has dimension $2^{2n}$ over the base field.  
The Clifford algebra is filtered by assigning filtration degree $+1$ to each generator; its associated graded algebra is the exterior algebra.

An explicit set of matrices realizing the even-dimensional Clifford algebras can be constructed iteratively, as follows (the method is due to Brauer and Weyl~\cite{BrauerWeyl}): Suppose generators $\Gamma_\mu^{(2n)}$ of~$\Cl_{2n}$ have already been constructed. Then generators 
$\Gamma_\mu^{(2n+2)}$ for~$\Cl_{2n+2}$ are given by 
\begin{equation}
\Gamma_\mu^{(2n+2)} = \Gamma_\mu^{(2n)} \otimes \sigma_3, \quad
 \Gamma_{2n+1}^{(2n+2)} = 1 \otimes \sigma_1, \quad
 \Gamma_{2n+2}^{(2n+2)} = 1 \otimes \sigma_2.
 \label{eq:BrauerWeyl}
 \end{equation}
These are $2n$ square matrices of size $2^n$. After remembering that we can use the defining anticommutation relation to antisymmetrize, there are $2^{2n}$ independent products of such gamma matrices, which additively span the bispinor. They are also invariant tensors, providing a complete set of intertwiners witnessing the relation
\deq[eq:bispinor]{
D\otimes D \cong \wedge^* V.
} We will denote such an antisymmetrized product by $\Gamma^I$, where $I\subseteq \{0,\ldots,d-1\}$.
 
 An additional matrix can be constructed, anticommuting with all of the gamma matrices: it defines a notion of chirality on spinors. In fact, it can be thought of as arising from dimensional reduction from the $(2n+1)$-dimensional Clifford algebra:
 \begin{equation}
 \overline\Gamma = (-i)^n \Gamma_1\cdots\Gamma_{2n}.
 \end{equation}
 It is clear from the explicit construction~\eqref{eq:BrauerWeyl} that
\deq[eq:chirality]{
\overline\Gamma = \sigma_3 \otimes \cdots \otimes \sigma_3.
}
There are corresponding projection operators onto the $S_\pm$ representations, which take the form
\begin{equation}
\pi_\pm = \frac{1}{2} \qty(1\pm \overline\Gamma).
\end{equation}
Since the $\Gamma$ matrices commute with $\overline\Gamma$, we can (if we like) write a pair of ``Weyl'' gamma matrices, $\gamma_\mu$, which carry indices for the $S_\pm$ representation rather than for the $D$. We will often simply write these as $\Gamma$; a contraction with a Weyl spinor shows which interpretation is intended.

We can decompose the invariant tensors~\eqref{eq:bispinor} in terms of their parity; obviously, those with odd numbers of indices behave like $\Gamma$ itself, while those with even numbers of indices are reversed. In ten dimensions, this leads to the useful identities
\begin{align}
\Sym^2 S_{+} & \cong V \oplus (\wedge^5 V)_{+}, \nonumber \\
\wedge^2 S_{+} & \cong \wedge^3 V, \\ 
S_{+} \otimes S_{-} & \cong \wedge^0 V \oplus \wedge^2 V \oplus \wedge^4 V. \nonumber
\end{align}
Each irreducible component is picked out by an intertwiner which is the gamma matrix with the appropriate number of indices. Here the $+$ subscript denotes a self-duality condition; see~\cite{Georgi} for details.

The space of degree-two elements in the Clifford algebra (or more precisely in its associated graded) is spanned by commutator elements 
\deq{
 M_{\mu\nu} = \frac{1}{4} [\Gamma_\mu, \Gamma_\nu ].
}
It is then not difficult to check the following identities by direct calculation:
\deq{
[M_{\mu\nu}, \Gamma_\rho ] = \Gamma_{\mu} \delta_{\nu\rho} - \Gamma_\nu \delta_{\mu\rho},
}
and further,
that
$[M_{\mu\nu}, M_{\lambda\rho}] $ reproduces precisely the commutation relations for the corresponding generators of $\so(N)$ above.
This says that $M_{\mu\nu}$ generate the Lie algebra, and furthermore that $\Gamma_\mu$ transforms in the standard vector representation with respect to that algebra (as suggested by the index it carries).

\subsection{Branching rules}

We will make heavy use throughout the paper of the branching of representations under the inclusion $U(1)\times SU(n) \subseteq \Spin(2n)$, and so it is worth recalling how these phenomena are understood. Consider the following combinations of gamma matrices of~$SO(2n)$:
\deq{
a^j = \frac{1}{2} \left( \Gamma_{2j} + i \Gamma_{2j+1} \right),
\qquad
a_j = \frac{1}{2} \left( \Gamma_{2j} - i \Gamma_{2j+1} \right).
}
It is easy to check that these form a set of $n$ fermionic creation and annihilation operators, i.e., that they satisfy the anticommutation relations
\deq{
\{ a^j, a^k \} = 0; \qquad \{ a^j, a_k \} = \delta^{j}_{k}. 
}
In our explicit presentation for gamma matrices, they look like
\deq[eq:aplus]{
a^j = 1 \otimes \cdots 1 \otimes \sigma_+ \otimes \sigma_3 \otimes \cdots \otimes \sigma_3,
}
with the raising operator occurring at the $j$-th place. The lowering operator $a_j$, of course, contains a lowering operator instead.

Implicit in this description is an identification of the Dirac spinor space $D \cong (\C^2)^{\otimes n}$, which can be thought of as writing the value of the spinor index ($0\leq\alpha< 2^n$) in binary. One must only adopt the somewhat peculiar convention that the basis vector of $\C^2$ annihilated by $\sigma_-$ is called 1, and the other is called 0. One can label basis vectors by the set $I\subseteq \{1,\ldots,n\}$ where the digit 1 occurs. It's then immediate from~\eqref{eq:aplus} that
\deq[eq:sign-rule]{
a_{i_\ell}\cdots a_{i_1} \ket{\emptyset} = \epsilon_{i_1\ldots i_\ell} \ket{ I },
}
and similarly immediate from~\eqref{eq:chirality} that
\deq[eq:parity]{
\overline{\Gamma} \ket{I} = {\ms}^{\abs{I}} \ket{I}.
}

The key equation for branching rules is
\deq[eq:branch]{
a^i a_j - \frac{1}{2}\delta^i_{j} = \frac{1}{2} \qty[  M_{2i,2j} + M_{2i+1,2j+1} +i \qty( M_{2i+1,2j} - M_{2i,2j+1})],
}
which is obtained by a simple direct calculation.
Recalling the antisymmetry of the $M_{\mu\nu}$, this then gives an explicit map (by contracting indices $i$ and~$j$) from the space of anti-Hermitian matrices (i.e., the Lie algebra $\lie{u}(n)$) to the real span of the $M_{\mu\nu}$, i.e., to~$\so(2n)$. A generator $(T^a)^i_j$ is mapped to $a^j(T^a)^i_j a_i$.

The decomposition of the spinor representations of~$\so(2n)$ into irreducibles of $\lie{u}(n)$ is then immediate to see. The Dirac spinor representation is the fermionic Fock space generated by the $a^j$, on which $\lie{u}(n)$ acts so that the $j$ indices label the fundamental representation of $\su(n)$. With respect to the trace part, \eqref{eq:branch} shows that a state which is annihilated by all lowering operators (the vacuum) must have charge $-n/2$; each raising operator carries charge $+1$.

If $\square$ is the fundamental representation of~$\su(N)$, then we've made an identification of the Dirac spinor with $\wedge^*(\square)$.
Also, since the $a^j$ are linear expressions in the $\Gamma$ matrices, they anticommute with $\overline\Gamma$. Therefore, the decomposition of $S$ into $S_\pm$ corresponds to the decomposition
\deq{
\wedge^*(\square) = \wedge^\text{odd}(\square) \oplus \wedge^\text{even} (\square).
}
The parity corresponding to $S_+$ depends on the dimension: for $\so(2n)$, the positive-chirality representation $S_+$ consists of the parity-$\ms^n$ tensors. (One remembers this by noticing that the vacuum state, annihilated by all $a_j$'s, carries an index whose binary expansion is all ones; as noted in~\eqref{eq:parity}, $\overline\Gamma$ counts the number of ones in the binary expansion, modulo two.) In ten dimensions, therefore, $S_+ = \wedge^\text{odd}(\square)$, and we have the decomposition 
\deq[eq:holspin]{
S_+ \rightarrow \rep{5}^{-3/2} \oplus \crep{10}^{1/2} \oplus \rep{1}^{5/2}, \qquad
S_- \rightarrow \rep{1}^{-5/2} \oplus \rep{10}^{-1/2} \oplus \crep{5}^{3/2}
}
under $ SU(5)\times U(1)\subseteq\Spin(10)$. The vector representation, of course, decomposes as 
\deq[eq:holvect]{
\rep{10} \rightarrow \rep{5}^{+1} + \crep{5}^{-1},
}
and it is not difficult to see that the two-form $\rep{45} = \wedge^2 (\rep{10})$ branches to 
\deq[eq:holadj]{
\rep{45} \rightarrow \rep{1}^{0} \oplus \rep{10}^2 \oplus \crep{10}^{-2} \oplus \rep{24}^0, \qquad
F_{\mu\nu} \mapsto    \{F=F_j^j,  F^{jk}, F_{jk}, F_j^k  \} .
}
The $\rep{24}$ is, of course, the adjoint representation of~$SU(5)$.

We'll also require the decomposition of the ten-dimensional gamma matrices themselves into $SU(5)$ invariant tensors. Luckily, this is now trivial to compute, since the decomposition of the vector index of $\Gamma$ just reduces it to the pair of~$a^j$ and $a_j$! We can replace ``slashed'' expressions as follows:
\deq{
\Gamma^\mu p_\mu \rightarrow a^j p_j + a_j p^j.
}
Now, the gamma matrices are invariant tensors for $SO(2n)$, and so (after reduction) they must become invariant tensors for $SU(n)$. There are two such basic tensors: the contraction of fundamental and antifundamental indices, $\delta^i_j$, and the Levi--Civita symbol $\epsilon_{jklmn}$ (with either raised or lowered indices). The only things to fix are signs. 

\section{Antifields}
\label{sec:BV}

We briefly recall some basic principles of the antifield formalism for quantization of gauge theories, due originally to Batalin and Vilkovisky. For a more thorough exposition, the reader is referred to the classic papers~\cite{BV,Schwarz-BV}, as well as to the recent review~\cite{Fiorenza}; of course, none of the material here is original, and we are heavily indebted to these and other sources in our presentation. However, we take the liberty of recalling these notions with some care, because they are important in what follows; we hope that this discussion serves to keep the reader properly oriented.

Throughout this section, we will let $\phi^i$ denote a collection of fields, which may be bosons or fermions and carry arbitrary Lorentz and other indices. $\phi^i$ are thought of as coordinates on a space~$V$, which is generally a superspace: its algebra of functions $\Fun(V)$ is $\Z/2\Z$-graded, and is commutative with respect to the Koszul sign rule.  

\subsection{An analogy}
\label{sec:analogy}

The intuitive picture of the BV procedure to keep in mind is the following. For the time being, let $V$ be an ordinary real vector space. Then $T^*V\cong V\oplus V^\dual$ carries a canonical symplectic structure, which can be written $\omega = d\phi^i \wedge d\phi_i^\dual$ with respect to a set of linear coordinates on~$V$. $\Fun(V)$ is therefore equipped with a Poisson bracket.

Suppose we wish to evaluate an integral of the form
\deq{
\int_V \alpha,
}
where~$\alpha$ is a top form on~$V$. We can attempt the following procedure: Extend $\alpha$ to a form $\tilde\alpha$ on~$V\oplus V^\dual$, satisfying the conditions 
\deq{
\left. \tilde\alpha \right|_V = \alpha, \qquad
d\tilde\alpha = 0.}
If we choose any function $f \in \Fun(V)$, then $\graph(df) \subset V\oplus V^\dual$ is a Lagrangian submanifold with respect to~$\omega$. We can then solve for the dual variables by setting 
$\phi_i^\dual = \pdv*{f}{\phi^i}$, and integrate the result, obtaining a result equivalent to the one we originally wanted. (One can imagine contour integration as a procedure of this kind.) Of course, the Lagrangian condition is not strictly necessary here, since by construction the integral of $\tilde\alpha$ depends only on the homology class of the cycle. However, in the BV formalism (which is a graded analogue of the procedure just discussed), this condition will be important.

This intuitive picture will be complicated by (at least) two additional subtleties. The first is that everything must be thought of in a graded-commutative, rather than commutative, setting, and there are some degree shifts. The second is that the original action may contain gauge fields, and therefore we may need to account for taking the invariants of a Lie algebra. For that, we use the Koszul complex (or BRST procedure), which will be reviewed quickly in the next section.

\subsection{The BRST differential}
\begin{satz}
The data of a Lie algebra structure on a vector space~$\lie{g}$ is equivalent to the specification of a degree-one differential $\kappa$ on~$\Fun(\lie{g}[1]) \cong \wedge^* \qty( \lie{g}^\dual )$. 
\end{satz}

Let us explain this in some detail. To define 
$\kappa$ starting with the Lie bracket $[,]:\wedge^2 \lie{g} \rightarrow \lie{g}$, one takes the dual map, which is naturally of the form $\lie{g}^\dual \rightarrow \wedge^2 \qty( \lie{g}^\dual ) $, and extends it to functions of all degrees by requiring the Leibniz rule. As the reader can easily check, nilpotence of~$\kappa$ is then equivalent to the Jacobi identity for the bracket. The inverse procedure is obvious. (If we were to weaken the requirement that $\kappa$ is of degree one, we would obtain the structure of an $L_\infty$ algebra.) 

Let $c^a$ be a set of functions dual to a choice of basis in~$\lie{g}$. Then the coordinate expression for the differential is 
\deq{
\kappa = \frac{1}{2} f^a_{bc} c^b c^c \pdv{}{c^a},
}
where $f^a_{bc}$ are the structure constants of the Lie algebra. The cohomology of the complex~$\qty(\Fun(\lie{g}[1]) ,\kappa)$ is (perhaps by definition) the Lie algebra cohomology of~$\lie{g}$.

Now, we will also be interested in representations of Lie algebras. Recall that 
the data of a module structure on a vector space $V$ is a set of representation matrices $(\rho_a)^i_j$, where $\phi^i$ is a set of coordinates on~$V$; more invariantly, it is a map
\deq{
\rho: V \otimes \lie{g} \rightarrow V.
}
\begin{satz}
\label{satz:Koszul}
Let  $\lie{g}$ be a Lie algebra and  $V$ a $\lie{g}$-module. The bracket structure and the module data $\rho$ are together equivalent to specifying a degree-one differential $\kappa$ on~$\Fun(V\oplus\lie{g}[1])$. Note that $V$ is not shifted in degree; we also insist that $\kappa$ has polynomial degree one.
\end{satz}
Again, the proof is a matter of unpacking definitions. Recall that
\deq{
S^2(V\oplus\lie{g}[1]) \cong S^2(V) \oplus \qty(V\otimes\lie{g}[1]) \oplus S^2(\lie{g}[1]).
}
Of course, since the Lie algebra is in odd degree, the last summand is really $\wedge^2(\lie{g})$. Just as the bracket was a map from this space to~$V\oplus \lie{g}$ in the case $V=\text{pt.}$, we can consider the map defined by $\rho + [,]$. The dual to this map defines a degree-one map on linear functions inside $\Fun(V\oplus\lie{g}[1])$, which can be extended uniquely as a derivation $\kappa$ to the whole space of functions. The condition that $\kappa^2=0$ is equivalent to requiring that $\lie{g}$ be a Lie algebra and that $V$ be a representation. In coordinates, we can write the differential in the form
\deq{
\kappa = (\rho_a)^i_j \phi^j c^a \pdv{}{\phi^i} + \frac{1}{2} f^a_{bc} c^b c^c \pdv{}{c^a}.
}
$\kappa$ is the BRST differential. Its cohomology in degree zero computes the $\lie{g}$-invariants inside of~$\Fun(V)$. In math, the pair $\qty( \Fun(V\oplus\lie{g}[1]), \kappa)$ is known as the \emph{Koszul complex}. We have recovered the statement that the physical observables of a gauge theory (i.e. its collection of gauge-invariant expressions whose expectation values may sensibly be computed) correspond to the BRST cohomology at ghost number zero; one can simply think that the $\phi^i$ are the fields of the theory, and the $c^a$ its ghosts.

The reader having experience with physical discussions of the BRST procedure may remark that the usual discussion typically involves additional new fields, often called $\bar{c}$ and~$b$, introduced as part of the Faddeev--Popov gauge fixing. We will comment on the origin of such fields in the antifield formalism below. 

Lastly, we note that both of these remarks apply equally well to super Lie algebras. The definitions must only be modified to take note of the additional $\Z/2\Z$ grading; overall parity is then determined by the sum of the Chevalley--Eilenberg degree (or ghost number) and the fermion parity (or superdegree). The differentials defining the structure then must carry ghost number one and superdegree zero, as well as polynomial degree one. Checking this is left as a pleasant exercise to the reader.

\subsection{The antifield procedure}

The antifield procedure follows closely the idea shown before in~\S\ref{sec:analogy}, except that there is an overall shift of parity. Concretely, for every field $\phi^i$ in the theory, one introduces a new \emph{antifield}, $\phi_i^\af$, with parity opposite to that of $\phi^i$. (If the space of fields was $W$, we are instructed to consider the new space $W\oplus W^\dual[-1]$.)

Just as $V\oplus V^\dual$ carried a canonical symplectic structure, so $W\oplus W^\dual[-1]$ carries a canonical \emph{odd} symplectic structure, called the \emph{antibracket:}
\deq[eq:antibracket]{
\antibr{ f, g } = \dR{f}{\phi^i} \dL{g}{\phi_i^\af} + \dR{f}{\phi_i^\af}\dL{g}{\phi^i}.
}
The antibracket clearly carries degree $+1$. 
One can quickly check that it also has the (graded) symmetry property
 \deq{
 \antibr{ f,g } = - \ms^{(\abs{f}+1)(\abs{g}+1)} \antibr{ g,  f }
 }

There is another canonical degree-one operator, the \emph{Batalin--Vilkovisky Laplacian,} which naturally acts on $\Fun(W\oplus W^\dual[-1])$. It is defined by the rule
\deq{
\Delta = \pdv{}{\phi_i^\af} \pdv{}{\phi^i}
}
One can straightforwardly check that this operator is nilpotent. However, it does not obey a Leibniz rule: in fact, 
\deq[eq:failure]{
\Delta (fg) = (\Delta f) g + \ms^{\abs{f}} f (\Delta g) + \ms^{\abs{f}} \antibr{ f, g } .
}
This is a good way to remember the definition~\eqref{eq:antibracket}.
The antibracket thus measures the failure of $\Delta$ to be a derivation, with respect to the ordinary product of functions. As a consequence, it is not necessarily true (unlike in a typical cdga) that the product of a closed and an exact object is itself exact! This fact is the source of much of the confusion in the study of BV cohomology, and we will remark further on it in what follows.

On the other hand, the adjoint operation of the bracket, $\ad_f(\cdot) = \antibr{ f, \cdot }$, is a (graded) derivation with respect to both the product and the bracket. That is, it satisfies analogues of both the Poisson and Jacobi identities:
\begin{align}
\ad_f (gh) &= \ad_f (g) \cdot h + \ms^{\abs{g}(\abs{f}+1)} g \cdot \ad_f(h), \\
\ad_f \antibr{g,h} &= \antibr{ \ad_f(g), h } + \ms^{(\abs{g} + 1 )( \abs{f}+1 )} \antibr{ g, \ad_f(h) } .
\end{align}
Furthermore, the BV-Laplacian \emph{is} a derivation with respect to the bracket:
\deq{
\Delta\{f,g\} = \antibr{ \Delta f, g } + \ms^{\abs{f}+1} \antibr{ f, \Delta g }.
}

In fact, as~\eqref{eq:failure} clearly shows, the antibracket is not an independent piece of structure: it can be recovered from the ordinary product together with the BV Laplacian. However, a BV algebra is slightly more than a cdga in which the condition that the differential be a derivation has been removed; 
the Jacobi identity for the antibracket translates into a seven-term relation expressing a constraint on the interaction of the product and BV Laplacian. We refer the interested reader to~\cite{Fiorenza}.

\begin{satz}[Main lemma~\cite{Schwarz-BV}]
Integration over Lagrangian submanifolds descends to a pairing between the cohomology of the BV-Laplacian $\Delta$ and the homology of the submanifold.
\end{satz}

For the proof of this statement, we refer the interested reader to~\cite{Schwarz-BV}. Of course, what the homology class of a supermanifold means is perhaps unclear; really, one should refer to the homology class of the ``body'' of the object. However, we will not need such generality; for our purposes, it will be sufficient to consider Lagrangians analogous to those of the form $\graph(df)$ mentioned previously.

Choose any fermionic element $f\in \Fun(W)$. 
$f$ is often called the \emph{gauge-fixing fermion}. Then the manifold defined by the conditions
\deq[eq:gff]{
\phi_i^\af = \pdv{f}{\phi^i}
}
is (shifted) Lagrangian with respect to the antibracket. The main lemma then implies that we can make any choice of gauge-fixing fermion we like; the path integral of a $\Delta$-closed integrand over the resulting space of fields, after imposing~\eqref{eq:gff}, will not depend on the choice of~$f$, and will produce the correct Feynman rules for the theory whose fields are $W$. Note that, in the absence of any further assumptions, imposing~\eqref{eq:gff} may break an integer grading to a $\Z/2\Z$ grading; it must, however, preserve $(-)^F$.

The problem then becomes to find a class of $\Delta$-closed action functionals, $\BVS(\phi^i,\phi_i^\af)$, that lead to sensible Feynman integrals in this way, and to relate them to standard actions $S(\phi^i)$ for familiar quantum field theories.

\begin{satz}[Quantum master equation]
The integrand of the path integral, $\exp(-\BVS/\hbar)$, is $\Delta$-closed (and therefore defines a BV cohomology class) if and only if $\BVS$ satisfies the so-called quantum master equation,
\deq[eq:qmaster]{
\antibr{\BVS,\BVS }  = 2\hbar\, \Delta \BVS.
}
\begin{proof}
The key calculation is 
\deq{
\Delta(\BVS^n) = n\, \BVS^{n-1} \Delta \BVS + \frac{n(n-1)}{2}\BVS^{n-2} \antibr{\BVS,\BVS},
}
which can be proved by a simple induction on~$n$. Substituting this into the condition that the integrand be closed then demonstrates that
\begin{align}
 0 &= \Delta \qty( e^{-\BVS/\hbar} )   \nonumber \\
 &= \qty( -\frac{1}{\hbar} \Delta \BVS + \frac{1}{2\hbar^2} \antibr{\BVS,\BVS } ) e^{-\BVS/\hbar} ,
\end{align}
from which~\eqref{eq:qmaster} follows.
\end{proof}
\end{satz}

Note that the quantum master equation can be solved order by order in the parameter~$\hbar$: if we expand the BV  action in powers of this parameter, 
\deq{
\BVS = \sum_{k\geq 0} \hbar^k \BVS_k ,
}
then it is easy to collect terms. In particular, $\BVS_0$ must then satisfy the classical master equation $\antibr{\BVS_0,\BVS_0} = 0$. 

\begin{satz}[Observables in the antifield formalism]
The gauge-invariant observables of a BV theory are identified with the cohomology of a different operator, namely 
\deq{
\bvd = \ad_\BVS - \hbar\, \Delta,
}
which is a nilpotent differential acting on the space of fields.
\begin{proof}
This, too, follows from a simple computation. We know that an operator $\phi$ will define a good integrand if and only if the integrand $\phi\, \exp\qty(-\BVS/\hbar)$ of the path integral is $\Delta$-closed.
But, if the action is assumed to satisfy~\eqref{eq:qmaster}, then this reduces to the condition that $\bvd(\phi) = 0$. Likewise, when $\phi$ is $\bvd$-exact, the total integrand is $\Delta$-exact, and therefore the expectation value must vanish. The fact that $\bvd$ is square-zero is checked by noting that its square is proportional to the quantum master equation for~$\BVS$.
\end{proof}
\end{satz}
Note that $\bvd$ suffers from the same handicap as the Batalin--Vilkovisky Laplacian: it fails to be a derivation with respect to the product of fields (except when the fields in question have zero antibracket). The algebra structure on fields therefore does not trivially descend to the space of observables. However, if one is just interested in the observables generated by expressions that do not depend on antifields, $\Delta$ vanishes trivially, and so $\bvd$ restricts to a derivation on this subspace.

\subsection{Solving the master equation in gauge theory}

We are now tasked with putting together the results we have developed in the previous subsections. The main remaining questions are: what does the BRST formalism for gauge theory have to do with the antifield formalism? And how can one go about producing actions $\BVS$ that can be meaningfully studied within that formalism, i.e., solutions to the quantum master equation?

This brings us to the last key idea, which is the analogue of the ``extension problem'' in~\S\ref{sec:analogy}. We will demonstrate that there is a canonical way to ``complete'' the action $S(\phi^i)$ of a gauge theory to a BV action $\BVS_0(\phi^i, \phi_i^\af)$, linear in the antifields, and satisfying the classical master equation. This procedure can be inverted: any solution of the classical master equation that is linear in antifields defines a complex of the form given in Fact~\ref{satz:Koszul}, and therefore can be thought of as a gauge theory.
Furthermore, under certain simple criteria on the gauge theory, $\BVS_0$ in fact solves the \emph{quantum} master equation as well, and the procedure is complete.

\begin{satz}[Action principle]
Let $W=V\oplus \lie{g}[1]$, for~$V$ a $\lie{g}$-module, and let $\kappa$ be the BRST differential on $\Fun(W)$, as defined in Fact~\ref{satz:Koszul}. Let $S$ be a degree-zero element of~$\Fun(W)$, for which $\kappa \qty(S) = 0$. Then the action
\deq{
\BVS_0 = S +  \phi_i^\af \kappa \qty(\phi^i)
}
is a solution to the classical master equation. Moreover, the adjoint action of $\BVS_0$ reproduces the Chevalley--Eilenberg differential $\kappa$ on the fields and ghosts of the original theory.
\label{satz:action}
\end{satz}
It is important to note that all of these terms appear in $\hbar$-degree zero. 

Similar reasoning~\cite{Baulieu:1990uv, Brandt:1996uv} allows one to conclude that $\BVS_0$ may be constructed so as to preserve any global symmetry of the original Lagrangian, even when the algebra only acts on shell! To find the action of the symmetry on antifields (which is not \emph{a priori} meaningful), one includes ghosts and terms in the differential for the entire symmetry algebra of the theory---both gauge and global. The failure of the algebra to close off shell means that the BRST differential will fail to be nilpotent. However, general theorems generalizing Fact~\ref{satz:action} ensure that a solution to the classical master equation can be found, at the expense of including quadratic and possibly higher-order terms in antifields.
This differential encodes the structure of the symmetry transformations on all of the BV fields in the manner of Fact~\ref{satz:Koszul}.
We follow through this reasoning in detail in the case of ten-dimensional super Yang--Mills theory below.
(For a discussion of how anomalies are related to the quantum master equation, see~\cite{Troost}.)


Lastly, we should remark on the absence of the typical Faddeev--Popov fields $\overline{c}$ and~$b$. As one can see from the form of the standard BRST transformation, these form a trivial pair and do not contribute to cohomology; they can be included at will without altering the theory. The necessity of including them arises from the gauge-fixing process, in which we must build a gauge-fixing fermion $f(\phi)$ out of the original fields. With only the fields of the minimal BV complex, no such suitable choice is possible. But introducing a trivial pair and making a corresponding choice of gauge-fixing fermion recovers the usual Faddeev--Popov procedure precisely. We refer the reader to the literature~\cite{Fiorenza,BV} for details.

\section{Twisting}
\label{sec:twists}

\subsection{Taking invariants}


At root, to ``twist'' a theory means to take the invariants of a fermionic symmetry. This procedure can be performed whenever the chosen symmetry operator~$Q$ is nilpotent; the invariants are then the cohomology of that operator (which can be thought of as all invariants, modulo those that are invariant for an uninteresting reason, and belong to multiplets where $Q$ is in fact represented nontrivially). We will use the notation $\Nilp(d,\N)$ for the locus of nilpotent supercharges inside of the fermionic part of the algebra. Since nilpotence is a scale-invariant condition, this is a cone, and descends to the projectivization; the space of possible twists is thus $\Nilpp(d,\N)$.

In the context of supersymmetric gauge theory, one has already gone through the machinery of replacing  a local bosonic symmetry (the gauge transformations) by a global fermionic one (the BRST symmetry); one takes invariants by passing to the cohomology of the BRST differential. Taking the invariants of a global fermionic supersymmetry is thus entirely analogous. In this context, one thus often speaks of ``adding $Q$ to the BRST differential.'' It is worth remarking that that procedure produces the $E_\infty$ page of the spectral sequence of the associated bicomplex, whereas taking the $Q$-invariants of the gauge invariants would produce the $E_2$ page. Of course, these don't necessarily agree; however, we are  aware neither of any concrete example in which they fail to, nor of a theorem that guarantees collapse at~$E_2$.


Of course, taking the invariants of $Q$ does some violence to the theory: the symmetry algebra is broken to the commutant $Z(Q)$ of~$Q$. To be more precise, we mean the subalgebra 
\deq{
Z(Q) = \{ x \in A : [x,Q] \propto Q\}
}
of the super-Poincar\'e algebra $A$.
With respect to this algebra, $Q$ is tautologically a scalar, possibly carrying some $U(1)$ charges. In general, though, Lorentz generators will fail to commute with~$Q$ (since it is a spinor), and so (by definition) will $R$-symmetry generators. However, $Z(Q)$ may nonetheless contain a subalgebra, isomorphic to the Lorentz algebra, but defined by the graph of some nontrivial homomorphism $\phi$ from the Lorentz group to the $R$-symmetry group. In this case, Lorentz symmetry is unbroken, but only in this ``twisted'' form. This is the case of a standard topological twist; importantly, the twisting homomorphism $\phi$ is usually \emph{not} independent data, but is essentially fixed by the choice of~$Q$. (The exception is when $Z(Q)$ is in fact so large as to contain more than one distinct $\lie{so}(2n)$ subalgebra; this will happen, for example, in holomorphic twists of four-dimensional $\N=4$ theories.) 

More generally, the Lorentz algebra  will simply break to some subalgebra of~$Z(Q)$. In the case of a ``holomorphic'' twist, for example, this will be $U(n) \subseteq SO(2n)$. This means that there is an interesting stratification of~$\Nilpp(d,\N)$, by what type of twist a given operator generates. Topological twists (when they exist) will form the top stratum, whereas holomorphic twists occur in the strata of lowest dimension. One can thus take invariants of $Q$ even in a theory that does not admit a twisting homomorphism in the traditional sense, the price being that the resulting theory will not be topological, and will make sense only on some class of manifolds with appropriately reduced holonomy group. We study the varieties $\Nilpp(d,\N)$ and their stratification in detail in~\cite{ESW-twists}, from the perspective of twists; examples have been studied in detail in the preceding literature, for example in~\cite{MovshevSchwarz,Berkovits,Cederwall} and (for simple Lie superalgebras) in~\cite{Gruson, DS}. The work in the first set of references is in the context of the pure spinor formalism; the connection to twisting has not, as far as we know, explicitly been noted before.


\subsection{Nilpotent loci and pure spinors}

There are sixteen 
supercharges in 
ten-dimensional 
$\N=1$ supersymmetry, transforming in an $S_-$ representation of the Lorentz group~\cite{KapustinWitten}. 
One can construct an odd supersymmetry by taking any linear combination of them:
\deq{
 Q = \sum_\alpha u^\alpha Q_\alpha, 
 }
 where the $u^\alpha$ are arbitrary complex constants that transform in the $S_+$. 

Let us compute the nilpotent locus, $\Nilp(10,1)$. The supersymmetry algebra is 
\deq{
\{ Q_\alpha, Q_\beta \} = \gamma^I_{\alpha\beta} P_I.
}
So one trivially computes that $A$ is nilpotent precisely when
\deq[eq:PSconstraint]{
\{ Q, Q\} = \left( u^\alpha \gamma^I_{\alpha\beta} u^\beta \right) P_I = 0.
}
In other words, $Q$ is nilpotent precisely when the $u$ parameters satisfy the ``pure spinor'' constraint (i.e., the vector quantity in parentheses vanishes) inside of the representation space $S_+$. Note that the $u$ are not Grassmann; they are just complex numbers in the spin representation of the Lorentz group. This is obvious from their origin as parameters for the odd part of the supersymmetry algebra.

Note also that, in Lorentzian signature, the spinor representation is real, and so the gamma matrices would satisfy a Hermiticity condition. That would mean that~\eqref{eq:PSconstraint} would have no solutions for real $u$ (and in particular, none of the generators could be nilpotent). In Euclidean, the spinor is complex, and so there is no longer a Hermiticity constraint. But that also means that the form $u\Gamma u$ isn't definite anymore; the lines spanned by the two nilpotent generators, $Q_0$ and $Q_{31}$, lie inside $\Nilp(10,1)$.

The space of solutions to the pure-spinor constraint is eleven-dimensional; it is (as always) a complex cone over its projectivization, a space of ten complex dimensions: 
\deq{
\Nilp(10,1) \subset \C^{16}, \qquad
\Nilpp(10,1) \subset P^{15}(\C).
}
In fact, this space is precisely equal to the space of inequivalent complex structures on $\R^{10}$: namely, 
\deq{
 \Nilpp(10,1) \cong SO(10)/U(5). 
 }
Of course, these holomorphic twists are all in a certain sense equivalent, because an $SO(10)$ Lorentz transformation carries any chosen complex structure into any other. But it is perhaps more useful to think of $\Nilpp(10,1)$ as parameterizing the set of choices to be made as to how Lorentz symmetry is broken in a given frame. The stratification is trivial, in this case: the algebra admits only minimal (holomorphic) twists. 

The next question to ask is, what does the tangent space to the space of pure spinors at a point look like, inside of the space of all supercharges, in terms of the decomposition of spinors induced by the reduction of the Lorentz group corresponding to the chosen point? In other words, what are infinitesimal parameters $u^\alpha$ such that $Q + u^\alpha Q_\alpha$ is still nilpotent, at linear order in the $u$'s?

Choosing a nilpotent $Q$ fixes a reduction of the Lorentz group from $SO(10)$ to $U(5)$, corresponding to a choice of complex structure. 
We'll adopt the standard notation $Q,Q_m, {Q}^{mn}$ for the decomposition of the $S_-$ spinor, and $P^m, {P}_{m}$ for that of the vector. Working out what the commutation relations of the algebra reduce to, one finds
\deq{
\{Q, Q\} = 0, \quad \{Q, Q_j \} = P_j,
\quad
\{Q,{Q}^{jk} \} = 0,
}
\[
 \{Q_j, Q_k \} = 0, \quad
  \{ Q_j, {Q}^{kl} \} \sim \delta_j^k P^{l} - \delta_j^l P^k, \quad
  \{ {Q}^{j k}, {Q}^{l m} \} \sim \epsilon^{jklmn} P_n.
\]
So the tangent space (at least at linear order) consists of the $\mathbf{10}$ representation.


In fact, the pure spinor constraint can be solved as follows:
\deq[eq:PSC]{
u u^m + \epsilon^{mnpqr}u_{np}u_{qr} = 0.
}
After fixing $u=1$ and solving for~$u^m$, this gives coordinates on~$\Nilpp(10,1)$ in terms of the representation space of the~$\mathbf{10}$.
Note that there are actually two equations that follow from the nilpotence of~$u^\alpha Q_\alpha$, one from the $P_m$ and one from the $P^m$ component. However, the other equation follows from this one, and so~\eqref{eq:PSC} is enough to ensure nilpotence of the deformed supercharge even for finite perturbations~\cite{Cederwall:2011yp}.

Interestingly, nilpotence varieties carry a number of tautological bundles, which allow one to think about the family of possible twists of a theory in a global fashion. The most interesting of these is the complex line bundle of nilpotent operators itself, which (for a Lie superalgebra $A = A^0 \oplus A^1$) is a subbundle of the trivial $A$-bundle over~$Y(A)$:
\begin{equation}
\begin{tikzcd}[column sep = 1 ex]
\mathscr{N}   \arrow[hookrightarrow]{rr} \ar[dr]  & & {A\times Y(A)} \ar[ld] \\
& Y(A)  & 
\end{tikzcd}
\label{eq:nilpbun}
\end{equation}
A similar construction defines a sheaf of algebras whose stalk at a point $Q\in Y(A)$ is the idealizer subalgebra $Z(Q)$. This is not an honest vector bundle on~$Y(A)$, since the fiber dimension may jump at different strata. However, it defines a vector bundle (with additional algebraic structure) over each pure stratum.

By an associated-bundle construction, the bundle $\mathscr{N}\rightarrow Y$ makes any $A$-module $V$ into a bundle of chain complexes over~$Y$. In other words, we just tensor with  the coordinate ring $\mathscr{O}(Y)$, producing the space of (holomorphic) sections of the trivial $V$-bundle over~$Y$, and then act in this by the nilpotent differential $Q = u^\alpha Q_\alpha$ ($u$ being coordinates on~$Y$). More generally, we could take any bundle of $A$-modules over~$Y$ (for example, by tensoring $V$ with the space of sections of a nontrivial line bundle). For a theory whose fields are~$V$, doing this construction over a point in~$Y$ produces the corresponding choice of twist of~$Y$; doing it globally produces \emph{all} possible twists of the theory, as a natural family over~$Y$ (the moduli space of twists of the theory). 

We remark that nilpotence varieties, as mentioned before, have been featured prominently in the pure spinor superfield formalism. In our language, the content of this formalism is as follows: One is interested in a multiplet, like the vector multiplet in ten-dimensional super Yang--Mills theory, for which no auxiliary-field formalism exists. Superfields, of course, do exist (for example, the unconstrained function ring $C(\R^{10|16})$), but they are much too large. As we have emphasized, truncations of quantum field theories essentially arise by taking invariants of (bosonic or fermionic) symmetries, i.e., by the twisting procedure. So one simply applies the (global!) twist mentioned above to the unconstrained superfield, producing the complex
\deq{
\left( C(\R^{10|16}) \otimes \mathscr{O}\left[ \Nilpp(10,1) \right], \mathscr{D} = u^\alpha \mathscr{D}_\alpha\right).
}
Rather magically, this produces exactly the BV complex of (untwisted) ten-dimensional Yang--Mills theory; see~\cite{Cederwall} and references therein. In the so-called ``non-minimal'' pure spinor superfield formalism, one resolves the holomorphic functions on~$Y$ by replacing them by the Dolbeault complex $\left(\Omega^{0,*}(Y),\bar{\partial} \right)$. One remark is in order: Like any unconstrained superfield, $C(\R^{10|16})$ admits \emph{two} commuting actions of the super-Poincar\'e algebra, by operators typically called $Q$ and~$\mathscr{D}$. Here, we have chosen to twist by the $\mathscr{D}$-action; the $Q$-action then gives the action of supersymmetry on the twisted superfield. One can follow the procedure outlined here for other choices of line bundles on~$Y$ and other choices of superalgebra, producing various well-known supermultiplets.

We close with an interesting point of speculation, to which we look forward to returning in future work. In the pure spinor superfield formalism, one writes an action for the pure spinor superfield which is precisely of holomorphic Chern--Simons form; this is equivalent to the usual Yang--Mills action for the component fields. As we will review below, the holomorphic twist of Yang--Mills is also a holomorphic Chern--Simons theory~\cite{Baulieu:2010ch,Costello:2016mgj}. So it is natural to conjecture that the pure spinor superfield, upon this further twist by the physical supersymmetry, simply becomes the antiholomorphic form degree of freedom. Moreover, one might suspect that a similar story could play out in greater generality: can, for instance, the untwisted four-dimensional $\N=1$ vector multiplet be formulated in terms of an appropriate pure-spinor superfield and an action of $BF$ type? 

\subsection{Twisting an interacting theory: the $F$-term spectral sequence}

In light of the above discussion, it is clear how one should proceed in general: First, formulate the BV complex for the theory, including the ghost sector encoding global symmetries. For the untwisted theory, set the superfluous ghosts to zero; for the twisted theory, set the (bosonic) supersymmetry ghost to a point on the nilpotence variety. The resulting complex, correspondingly, represents the twisted or untwisted theory. For the twisted theory, regrading will be necessary so that the differential $d = Q + \bvd$ has appropriate homological degree. In this language, it is obvious that there is a spectral sequence from the untwisted to the twisted theory, which is just the spectral sequence of the overall bicomplex. (A review of spectral sequences in physical language is given in~\cite[\S1]{Gukov:2015gmm}.)

However, the differential $d$ consists of several terms, which could be grouped differently. Each such meaningful separation leads to a different spectral sequence. In particular, as was pointed out in~\cite{Gukov:2015gmm}, one can separate those terms $d_0$ that are present in the supersymmetry tranformations of the free theory from those (called $d_1$) that appear only in the interacting theory. 
In fact, relics of this spectral sequence can already be seen in classical computations of the elliptic genus for Landau--Ginzburg models~\cite{Witten:1993jg}.
More generally, one could consider any supersymmetry-preserving deformation of a twisted theory. 

In fact, this spectral sequence is also implicitly used in previous computations of the superconformal index, for instance in~\cite{Chang:2013fba, Eager:2012hx}. For example, in the computation of the superconformal index of $\N=4$ super Yang--Mills theory~\cite{Chang:2013fba}, one first considers letters satisfying the BPS bound with respect to their free-theory conformal dimensions. By formal Hodge theory, this is the same as looking at $d_0$-cohomology. One then takes the cohomology of a further ``supercharge,'' which is in fact $d_1$. (Again, one is tacitly using the collapse of the spectral sequence at the $E_2$ page.) 

In~\cite{Eager:2012hx}, it was demonstrated that the differential $d_1$ acting on the cohomology of~$d_0$ reproduces the Ginzburg dg algebra of a quiver corresponding to the $\N=1$ gauge theory. However, in that case, all of the terms could be matched just by looking at the supersymmetry operator $Q$. We will find that in two dimensions, the same story persists, but it is crucial to recognize that the relevant differential is $d_1$, which in fact contains terms from both $Q$ and~$\bvd$. 

\section{Maximally supersymmetric Yang--Mills}
\label{sec:MSYM}

Keeping the discussion above in mind, we choose to formulate ten-dimensional Yang--Mills theory in terms of a differential that represents both gauge and global symmetries. The result is not the BRST differential of the theory (in other words, we do not pass to its cohomology); however, the BRST differential can be obtained simply by setting ``ghosts'' for super-Poincar\'e symmetry to zero. This technique is far from new; it was applied to four-dimensional supersymmetry multiplets in~\cite{Baulieu:1990uv}, while general theorems about on-shell global symmetry algebras were proved in~\cite{Brandt:1996uv}. 
Our other choices of conventions parallel those used in~\cite{Murat,KapustinWitten}.

The action of ten-dimensional super Yang--Mills theory is
\deq[SYMaction]{
S = \int d^{10}x\, \Tr \qty( -\frac{1}{4} F_{\mu\nu} F^{\mu\nu} + \frac{i}{2} {\lambda} \Gamma^\mu D_\mu \lambda).
}
Note that, although we write $\Gamma$ in such expressions, all relevant spinors obey a Weyl condition; the field $\lambda$, in particular, transforms in the $S_+$ representation. Therefore $\Gamma$ can be thought of as replaced with $\gamma$ everywhere, with raised or lowered indices as appropriate to the context.
Covariant derivatives and field strengths are defined by
\deq{
D_\mu = \partial_\mu + g[A_\mu, \cdot], \qquad
F_{\mu\nu} = \frac{1}{g} [D_\mu, D_\nu] = \partial_\mu A_\nu - \partial_\nu A_\mu +  g [ A_\mu, A_\nu].
}
Note that we have adopted the convention that the Lie algebra of $\lie{u}(N)$ consists of antihermitian, rather than hermitian, matrices. This action is invariant under the global super-Poincar\'e algebra, which acts on the fields by the transformations
\deq[SYMsusy]{
\delta A_\mu = i \epar \Gamma_\mu \lambda, \qquad
\delta \lambda = \Gamma^{\mu\nu} F_{\mu\nu} \epar.
}
In~\eqref{SYMsusy}, \epar{} is a fermionic parameter for the supersymmetry transformation, which pairs with $Q$ and therefore transforms in the representation $S_+$. One should understand that 
\deq{
\delta_\epar = \{ \epar^\alpha Q_\alpha, \cdot \} 
}
for any field or operator. It is worth reminding the reader of two important subtleties, typical of supersymmetric gauge theories: the transformations~\eqref{SYMsusy} generate the super-Poincar\'e algebra only up to gauge transformations, and only modulo the equation of motion for the fermion field $\lambda$.

The gauge transformations of the fields are
\deq[SYMgauge]{
\delta A_\mu = D_\mu \xi, \qquad
\delta \lambda = g [\xi, \lambda] ,
}
where the gauge parameter $\xi$ is a $\lie{g}$-valued scalar field.

The whole symmetry algebra of the theory (both gauge and global) can therefore be conveniently packaged in terms of a single differential, acting on an appropriate space of fields and ghosts:
\begin{align}
\delta A_\mu &= D_\mu c + a^\nu \partial_\nu A_\mu + i \epar \Gamma_\mu \lambda, 
\nonumber
\\
\delta \lambda &= g\{c,\lambda\} + a^\nu \partial_\nu \lambda + \Gamma^{\mu\nu} F_{\mu\nu} \epar,
\nonumber
\\
\delta c &= \frac{g}{2}\{c,c\} + a^\nu \partial_\nu c - 2i \epar \Gamma^\mu \epar A_\mu, 
\label{Cdiff}
\\
\delta a^\mu &= 2 i \epar \Gamma^\mu \epar, 
\nonumber
\\
\delta \epar &= 0.
\nonumber
\end{align}
Here, the gauge parameter has become the usual anticommuting ghost field $c$; the other ghosts, $\epar$ and $a^\mu$, are commuting and anticommuting respectively, and (since they are merely stand-ins for the global supersymmetry algebra) do not depend on the spacetime coordinate. The BRST differential $\kappa$ will be obtained by setting these to zero. Alternatively, for the holomorphically twisted theory, we will set $a^\mu$ to zero and $\epar$ to a choice of point on the nilpotence variety.

One term may deserve comment: the last part of the $c$-ghost transformation in~\eqref{Cdiff} represents the fact that the commutator of two supersymmetries produces both a translation and a gauge transformation. (Since the differential is the dual of the bracket map, this means that the gauge ghost acquires a term proportional to the square of the supersymmetry ghost.)

Moreover, since the algebra only acts on the fields modulo the fermion equation of motion, the differential we have written down fails to be nilpotent; indeed, its square is proportional to that equation of motion. It is not too difficult to check that
\deq{
\delta^2 \lambda = 2 i  \epar (\epar \gamma^\mu D_\mu \lambda),
}
which vanishes on-shell due to the Dirac equation, and that otherwise $\delta^2=0$. Following~\cite{Baulieu:1990uv}, we therefore add a quadratic term
\deq{
S^{(2)} = 2i (\epar\lambda^+)^2
}
to the BV action (including all ghosts for global and gauge symmetries). In the terminology of~\cite{Baulieu:1990uv} (which performed essentially an identical computation for four-dimensional $\N=1$ supersymmetry), the system is of second rank, and no further terms need to be added. The result for the complete BV action, applying the prescription of Fact~\ref{satz:action} to the transformations \eqref{Cdiff}, is
\begin{multline}
\BVS = \tr \int d^{10}x\, \left[ -\frac{1}{4} F_{\mu\nu} F^{\mu\nu} + \frac{i}{2} \lambda \Gamma^\mu D_\mu \lambda 
- (A^\af)^\mu \left( D_\mu c + a^\nu \partial_\nu A_\mu + i \epar \Gamma_\mu \lambda \right) 
\right.
\\
\left.
 - \lambda^\af \left( g\{c,\lambda\} + a^\nu \partial_\nu \lambda + \Gamma^{\mu\nu} F_{\mu\nu} \epar \right)
  - c^\af \left( \frac{g}{2}\{c,c\} + a^\nu \partial_\nu c - 2i \epar \Gamma^\mu \epar A_\mu  \right) 
  + 2 i (\epar \lambda^+)^2 \right].
\label{SYM-BV-S}
\end{multline}
Setting global ghosts $a^\mu$ and~$\epar$ equal to zero returns the standard BV action, agreeing with standard references~\cite[for instance]{BerkovitsGomez}. On the other hand, the action of a physical supercharge $Q = u^\alpha Q_\alpha$ on a field can be obtained by taking the adjoint action of~$\BVS$ on that field, and then setting the $\epsilon$ supersymmetry ghosts to~$u$ and other ghosts to zero. Essentially, this prescription reduces to changing the supersymmetry transformation for the fermion as follows:
\deq{
\delta \lambda = \Gamma^{\mu\nu} F_{\mu\nu} \epar - 2 i (\epar \lambda^+) \epar.
}
This is now a nonlinear action on the BV complex. The supersymmetry variations, as well as BRST variations, for the fields and antifields are straightforward to obtain, and we sum them up in Table~\ref{tBRST}. In addition, the fields of the theory in Berkovits' pure spinor formalism, with their $\theta$ and BV degrees, are shown in Table \ref{Bfields}~\cite{BerkovitsGomez}.

\begin{table}[th]
\caption{Fields of ten-dimensional super Yang--Mills, with $\Spin(10)$ indices}
\begin{center}
\begin{tabular}{|c|c|c|c|c|}
\hline
BV: & $1$ & $0$ & $-1$ & $-2$  \\
\hline
\hline
$\theta^{\strut 0}$ & $c$ ($\rep{1}$) &  & & \\
\hline
$\theta^{\strut 1}$ & & $A_\mu$ ($\rep{10}$) & & \\
\hline
$\theta^{\strut 2}$ &  & $\lambda_{\alpha}$ ($S_+ = \rep{16}$)  & & \\
\hline
$\theta^{\strut 3}$ &  &  & $ \lambda_{\alpha}^+$ ($S_- = \crep{16}$) & \\
\hline
$\theta^{\strut 4}$ &  &  & $A_\mu^+$ ($\rep{10}$) & \\
\hline
$\theta^{\strut 5}$ &  &  & & $c^+$ ($\rep{1}$) \\
\hline
\hline
\end{tabular}
\end{center}
\label{Bfields}
\end{table}

\begin{table}[th]
\caption{BRST and supersymmetry transformations}
\begin{center}
\begin{tabular}{|c|c|c|}
\hline
$\phi$ & $\kappa\phi$    &  $\{ u^\alpha Q_{\alpha}, \phi\} $ \\
\hline
\hline
$c$ & $\frac{g}{2} \{c,c\}$ & $0$
\\ \hline
$A_\mu$ & $D_\mu c$ & $u \Gamma^{\mu} \lambda$ 
\\ \hline
$\lambda $ & $- g \{c, \lambda\}$  & $\Gamma^{\mu\nu} F_{\mu\nu} u - 2 i u (u\lambda^+)$   
\\ \hline
$\lambda^\af$ & $ i \Gamma^\mu D_\mu \lambda +  g [ c, \lambda^\af]$ 
	& $- i A^{\af}_\mu \Gamma^{\mu} u $ 
\\ \hline
$A_\mu^\af$ & $ - D^\nu F_{\nu\mu} + i \lambda \Gamma_\mu \lambda + i \{ A_\mu^\af, c\}$
	 & $ u \Gamma^{\mu\nu}D_\nu \lambda^+$
\\ \hline
$(c^\af)_a$ & $ -D^\mu (A^\af_\mu)_a - g f^b_{ac} (\lambda^\af)_b \lambda^c
+ g f_{ac}^b   c^c c^\af_b $ & $ 0 $
\\ 
\hline
\hline
\end{tabular}
\end{center}
\label{tBRST}
\end{table}

\section{Holomorphic twist in ten dimensions}
\label{sec:holo}

\subsection{Reduction of the structure group}
In order to perform the holomorphic twist, one chooses a nilpotent element $Q = u^\alpha Q_\alpha$, or equivalently a point on the pure spinor variety $\Nilpp(10,1)\cong SO(10)/U(5)$. The choice is thus equivalent to a choice of complex structure on~$\R^{10}$, and we can simply branch all $SO(10)$ representations to the chosen $U(5)$ subgroup, using the rules given in~\S\ref{sec:spinors}.
After applying~\eqref{eq:holvect}, the supersymmetry transformation rules for the vector representation reduce to
\deq{
\delta A_m = i \qty( \epar^n \lambda_{nm} + \epar_{mn} \lambda^n ) , 
\qquad
\delta A^m = i \qty( \epar \lambda^m +  \epar^m \lambda + \epsilon^{mnpqr} \epar_{np} \lambda_{qr} ).
}
For the supersymmetry action on the spinor, one needs to work out the antisymmetrized product of two gamma matrices, considered as a map from~$S_+$ to~$S_+$. The spinor representation itself decomposes according to~\eqref{eq:holspin}, and the antisymmetric tensor $\rep{45}$ of~$SO(10)$ according to~\eqref{eq:holadj}. 
The result is
\begin{align}
\delta \lambda &= \epar F + \epar_{jk} F^{jk},
\nonumber \\
\delta \lambda^j &= \epar^j F + \epar^k F_k^j + \epsilon^{jklmn} \epar_{kl} F_{mn},
\\
\delta \lambda_{jk} &= \epar F_{jk} + \epar_{jk} F + \epsilon_{jklmn} \epar^l F^{mn}
\nonumber
\end{align}
\begin{table}[th]
\caption{Holomorphic decomposition}
\begin{center}
\begin{tabular}{|c|c|c|c|c|}
\hline
~BV:~ & ~$1$~ & ~$0$~ & ~$-1$~ & ~$-2$~  \\
\hline
\hline
$\theta^{\strut 0}$ & ~$\rep{1}^0$~ &  & & \\
\hline
$\theta^1$ & & ~$\rep{5}^{1} \oplus \crep{5}^{-1}$~ & & \\
\hline
$\theta^2$ &  & ~$\rep{5}^{-3/2} + \crep{10}^{1/2} + \rep{1}^{5/2} $~  & & \\
\hline
$\theta^3$ &  &  & ~$\crep{5}^{3/2} + \rep{10}^{-1/2} + \rep{1}^{-5/2} $~ & \\
\hline
$\theta^4$ &  &  & ~$\rep{5}^{1} \oplus \crep{5}^{-1}$~ & \\
\hline
$\theta^{\strut 5}$ &  &  & & ~$\rep{1}^0$~ \\
\hline
\hline
\end{tabular}
\end{center}
\label{tfields}
\end{table}
For convenience, we summarize the $U(5)$-invariant decomposition of the fields in Table~\ref{tfields}, and record the supersymmetry and BV transformations in holomorphic language in Table~\ref{rBRST}. In addition, one can write the (physical) BV action in holomorphic language:\begin{multline}
\BVS = \int d^{10}x\, \tr \Big( -\frac{1}{4} \qty( F^2 + F_m^n F^m_n + 2 F_{mn} F^{mn})  \\ 
+ i \qty( \lambda D_m \lambda^m + \lambda^m D^n \lambda_{mn} + \epsilon^{mnpqr} \lambda_{mn} D_p \lambda_{qr} )
+ i A^\af_m D^m c + i (A^\af)^m D_m c  
\\
- g \qty(  \lambda^\af c \lambda +  \lambda^\af_m c \lambda^m + ( \lambda^\af )^{mn} c \lambda_{mn} ) - g c c c^\af \Big).
\label{SYM-BV-holo}
\end{multline}
Note that this is identical to~\eqref{SYM-BV-S}. So far, all we have done is expressed the index structure differently. Of course, when we actually perform the twist of the theory by adding $Q$ to the BRST differential $\kappa$, some terms in this action will become exact and can be discarded. We will examine the effects of this twist in the next section.
\begin{table}[th]
\caption{BRST and scalar SUSY transformations in holomorphic language. Color is not shown, but can be restored by simply including commutators or anticommutators as appropriate.}
\begin{center}
\begin{tabular}{|c|c|c|}
\hline
$\phi$ & $\kappa\phi$ & $Q\phi$ \\
\hline
\hline
$c$ & $\frac{g}{2} c c$ & $0$
\\ \hline
$A_m, \ A^m$ & $D_m c,  \ D^m c$  & $0, \ \lambda^m$
\\ \hline
$\lambda,  \lambda^m, \lambda_{mn} $ & $- g c\lambda, \ - g c\lambda^m, \ - g c\lambda_{mn}$  
& $F-2i\lambda^+, \  0, \ F_{mn}$
\\ \hline
$\lambda^\af$ & $ i D_m \lambda^m +  g c \lambda^\af $ & $0$
\\ \hline
$\lambda^\af_m$ & $ i  D_m \lambda  + i D^n \lambda_{nm} +  g  c \lambda^\af_m $ & $A^\af_m$
\\ \hline
$(\lambda^\af)^{mn}$ & $ i D^m \lambda^n + i \epsilon^{mnpqr} D_p \lambda_{qr} +  g  c (\lambda^\af)^{mn}$ & $0$
\\ \hline
$A_m^\af$ & $ - D_m F - D_n F^n_m - D^n F_{nm}
+ ig \lambda^n  \lambda_{mn} + i g A_m^\af c$ & $0$
\\ \hline
$(A^\af)^m$ & $ - D^m F - D^n F^m_n - D_n F^{nm} 
+ ig \lambda  \lambda^m + i g \epsilon^{mnpqr} \lambda_{np} \lambda_{qr} + i g( A^\af )^m c$ & 
$D^m \lambda^\af + D_n (\lambda^\af)^{nm}$
\\ \hline
$c^\af$ & $ -D^m A^\af_m - D_m (A^\af)^m - g \lambda^\af \lambda + g  c c^\af $ & $0$
\\ 
\hline
\hline
\end{tabular}
\end{center}
\label{rBRST}
\end{table}

\subsection{Broken symmetries imply regrading}
Before $Q$ is added to the BRST differential, the $U(5)$-equivariant complex of fields is graded by homological degree, or ghost number, as well as by the $U(1)$ Lorentz symmetry which is the phase part of~$U(5)$. The new differential explicitly breaks a combination of these gradings: $Q$ has Lorentz grading $-5/2$, and ghost number zero, while $\kappa$ has ghost number one and zero Lorentz grading. The combination which is preserved is therefore 
\deq{
a = d_\text{L} - \frac{5}{2} d_{\text{BV}},
}
and the complex representing the twisted theory is graded by this quantum number. Of course, the differential $Q + \kappa$ sits in $a$-degree $-5/2$. It is straightforward to check the value of this quantum number for all fields, using Table~\ref{tfields}.
We emphasize that this regrading is not a choice; rather, it is forced on us by the manner in which the choice of $Q$ breaks the symmetries of the physical theory.

Using Berkovits' $\theta$-grading, another grading on the complex can be defined. We can, for example, form the combination 
\deq{
b =  \frac{3}{2}d_{\theta} - a.
}
The combination is motivated by the fact that all of the fields that survive the twist appear in uniform $b$-degree. We collect the $a$ and~$b$ gradings of all fields in Table~\ref{RegradedFields}. (It may be instructive to compare with~\cite[eq.~(4.14)]{Berkovits}.) Notice that we do not claim that the $b$-grading is unbroken by the twist! It is included as a way of making the table more legible; only the $a$-grading, which defines the relevant notion of homological degree in the twisted theory, will play a substantial role in what follows.\nfoot{In particular, it is not obvious how to assign $\theta$-degree to $Q$, $\kappa$, and derivative operators. The most plausible assignment places $Q$ in degree one and $\kappa$ in degree zero; if this is true, then $b$ is broken in the twisted theory, and only a different combination, $b' = d_\text{BV} + d_\theta$, will survive. Using such a combination, in fact, makes the cancellation between BRST-trivial pairs more apparent. However, we do not use the $\theta$ grading in what follows, and so defer these questions to future work.}
 The $a$-grading of the differential $Q+\kappa$ is $-5/2$; $(A^m,\lambda^m)$ and $(\lambda^+_m,A^+_m)$ form two trivial pairs and disappear from the cohomology. Similarly, the $\lambda$ and~$\lambda^+$ singlets cancel together with singlet components of the field strength~\cite{Baulieu:2010ch}. Finally, the derivative operators $\partial^m$ and~$\partial_m$ sit in $a$-degree $1$ and~$-1$, respectively.
\begin{table}[th]
\caption{Regraded table of fields}
\begin{center}
\begin{tabular}{|c|c|c|c|c|c|c|c|c|c|c|c|c|c|c|c|c|}
\hline
$a$ & $ -\frac{5}{2}$ & \,$-2$\, &  $ -\frac{3}{2}$ &  $-1$ &\, $ -\frac{1}{2}$ \,& $0$ & $\frac{1}{2}$ &  $1$ & $\frac{3}{2}$ & $2$ & ~$\frac{5}{2}$~ & ~$3$~ & $\frac{7}{2}$ & $4 $ & ~$\frac{9}{2}$~ & $5$    \\
\hline
\hline
$b=1/2$& &&& &&& & $A^m$ &&  &\,$\lambda$\,&& &$\lambda^+_m$&& \\
$b=5/2$ & $c$ &&& \,$A_m$\, &&&  \,$\lambda_{mn}$\, &&& $(\lambda^+)^{mn}$ &&&  $(A^+)^m$ &&& \,$c^+$\, \\
$b=9/2$ & && $\lambda^m$ & &&$\lambda^+$& &&\,$A^+_m$\,& &&& &&&  \\
\hline
\hline
\end{tabular}
\end{center}
\label{RegradedFields}
\end{table}

We can sum up the preceding discussion with the following nice picture: The holomorphic twist of ten-dimensional super Yang--Mills theory is holomorphic Chern--Simons theory. Its BV complex consists of a single form superfield, $\alpha \in \Omega^{0,*}(\C^5)$, with coefficients in the Lie algebra of the gauge group. For brane worldvolume theories, this will be~$\lie{u}(N)$. 

It is important to remark that the holomorphy of~$\alpha$ applies only to its form indices; the functional dependence is that of any smooth function on~$\R^{10}$. Therefore, this Dolbeault-like complex is a resolution of the sheaf of all holomorphic functions, rather than simply of the constant sheaf. Finally, the differential on the BV complex is given by the simple formula
\deq{
d\alpha = \bar{\partial}\alpha + g\alpha \wedge \alpha . 
}
(Again, this result was first found by Baulieu~\cite{Baulieu:2010ch}.) The reader will enjoy finding the corresponding terms in Table~\ref{rBRST}, and noting that they originate seemingly at random from the $\kappa$ and~$Q$ columns. In addition, the factors of $g$ have been included, so that it is apparent how to separate the free-theory differential $d_0$ from the interaction terms $d_1$: in particular, $d_0$ is just the $\bar\partial$ operator.
\section{Dimensional reductions and supersymmetric indices}
\label{sec:dimred}
Imagine a stack of $N$ D$(2k-1)$-branes, supported along $\C^k \subseteq \C^5$.
As we have seen in great detail, the field content of the holomorphic twist of the D$10$-brane woldvolume theory is described by holomorphic Chern--Simons theory on~$\C^5$, which can be thought of as the BV complex
\deq{
\left[ \Omega^{0,*}(\C^5) \otimes \lie{u}(N); d: \alpha \mapsto \bar{\partial} \alpha + g \alpha \wedge \alpha \right].
}
  The dimensional reduction of this theory was studied in~\cite{Costello:2016mgj}. One simply applies the standard procedure: some of the  $U(5)$ Lorentz indices become $R$-symmetry indices, as usual. The resulting theory is (the holomorphic twist of) maximally supersymmetric Yang--Mills on~$\R^{2k}\cong\C^k$, which (in the language of minimal supersymmetry) now contains both the vector multiplet and some additional matter. 
It is straightforward to see this in the language of~\S\ref{sec:holo}: the antiholomorphic form field $\alpha$ reduces to 
\deq{
\Omega^{0,*}(\C^k)\otimes \C[\psi_1,\ldots,\psi_{5-k}] \otimes \lie{u}(N),
}
where the new factor represents the odd generators $dz_i$ for directions transverse to the worldvolume. It is simply a (graded) polynomial algebra on odd generators $\psi_i$. This theory is the holomorphic twist of maximally supersymmetric Yang--Mills theory in $2k$ dimensions. 

One can view the exterior algebra $\C[\psi_1,\ldots,\psi_{5-k}]$ as the $\Ext$ algebra of the skyscraper sheaf, $\mathscr{O}_p$, supported at a point $p\in \C^{5-k}$ (which might as well be the origin in this case). Thus, one expects~\cite{Aspinwall-Katz} that the worldvolume theory of a D$(2k-1)$-brane probing a Calabi--Yau $(5-k)$-fold $X$ in the spacetime $\C^k\times X$ will have a holomorphic twist whose BV complex is
\deq{
\Omega^{0,*}(\C^k) \otimes \Ext^{\bullet}_{\mathscr{O}_X}(\mathscr{O}_p,\mathscr{O}_p).
}
In the case when $X$ is a Calabi--Yau cone, we should insist that $p$ is the (usually singular) point at the tip of the cone. In this paper, we will concentrate on the case of flat ten-dimensional spacetime. However, one should expect our results to hold for any Calabi--Yau compactification. Thus, the general setting in which one should imagine oneself (for example in two spacetime dimensions) is that of D$1$-branes in $\C\times X^4$; the worldvolume theory will then have $(0,2)$ supersymmetry, which can be used to perform the holomorphic twist.

Of course, we must also modify the form of the differential (equivalently, of the transformations in Table~\ref{rBRST}) to take into account the fact that some of the derivatives in ten dimensions do not survive dimensional reduction. Since derivatives are set to zero upon dimensional reduction, this just means replacing the $\bar\partial$ operator in five complex dimensions by its $k$-dimensional counterpart. 
This means that the collection of gauge-invariant local operators will differ as well: in ten dimensions, holomorphic Chern--Simons has no gauge-invariant local operators. As a prototypical example, one should think of ordinary abelian Chern--Simons in three dimensions. There, the BV complex is simply the de~Rham complex.  Since the de~Rham complex is a resolution of the constant sheaf, there are no local operators; Wilson lines are a complete set of gauge-invariant observables. Similarly in (standard) abelian holomorphic Chern--Simons, the BV complex is the Dolbeault complex, which has no cohomology in ghost number zero.  
Upon dimensional reduction, though, gauge invariant local operators will appear in the part of the theory which is in matter multiplets of minimal supersymmetry.
We study the cases $k=1,2$ in greater detail below.
\subsection{Dimensional reduction to four dimensions}
\label{sub:4d}
\begin{table}[thp]
\caption{Dimensional reduction of fields to four dimensions}
\[
\begin{tikzcd}[row sep = 1ex, column sep = 1ex]
(\rep{1},\rep{1}) & (\rep{1},\crep{3}) & (\rep{1},\rep{3}) & (\rep{1},\rep{1}) & & \\
        &(\rep{2},\rep{1}) & (\rep{2},\crep{3}) & (\rep{2},\rep{3}) & (\rep{2},\rep{1}) & \\
         & &  (\rep{1},\rep{1}) & (\rep{1},\crep{3}) & (\rep{1},\rep{3}) & (\rep{1},\rep{1}) \\
         \hline
\rep{1} & \crep{5} & \crep{10} & \rep{10} &  \rep{5} &  \rep{1}\\    
\end{tikzcd}
\]
\label{tab:4dred}
\end{table}
The ten-dimensional Lorentz symmetry is broken to a $SO(4)$ Lorentz group and an $SO(6)$ $R$-symmetry group upon dimensional reduction.  The structure group is reduced to $U(5)$ by the choice of a nilpotent supercharge.  Thus, in the twisted, dimensionally reduced theory, the relevant symmetry group (unbroken Lorentz and $R$-symmetry) is $U(2) \times U(3)$.  This is summarized in the following diagram of inclusions of groups:
\begin{equation}
\begin{tikzcd}
SO(10) & SO(4) \times SO(6) \ar[l] \\
U(5) \ar[u] & U(2) \times U(3) \ar[l] \ar[u]
\end{tikzcd}
\label{commsq}
\end{equation}
In diagrams such as these, the horizontal arrows are dimensional reduction, and the vertical arrows represent a minimal or holomorphic twist.
The branching rules for the vector representation are as follows:
\begin{equation}
\begin{tikzcd}
\mathbf{10} \ar[r] \ar[d] & (\mathbf{4},\rep{1})\oplus(\rep{1},\rep{6})\ar[d] \\
\mathbf{5}^{1}\oplus\smash{\crep{5}}^{-1} \ar[r] & (\mathbf{2}^1, \rep{1}^0) \oplus (\rep{1}^0, \mathbf{3}^{1}) \oplus (\mathbf{2}^{-1},\rep{1}^0) \oplus (\rep{1}^0,\crep{3}^{-1}).
\end{tikzcd}
\end{equation}
We see that the ten-dimensional vector decomposes into a four-dimensional vector and six scalars in the $\mathbf{6}$ of the $R$-symmetry group.  The holomorphic piece of the vector and the scalars in the $\crep{3}$ of~$U(3)$ survive in the holomorphic twist.

Similarly, the ten-dimensional spinor decomposes as:
\begin{equation}
\begin{tikzcd}
S_+ \ar[r] \ar[d] & (\mathbf{2}, \mathbf{4}) \oplus (\mathbf{\overline{2}} , \mathbf{\overline{4}}) \ar[d] \\
 \mathbf{5}^{-3/2} \oplus \smash{\mathbf{\overline{10}}}^{1/2} \oplus \mathbf{1}^{5/2} \ar[r] & 
 \qty(\rep{1}^{1} \oplus \rep{1}^{-1},\mathbf{1}^{3/2} \oplus \mathbf{3}^{-1/2} ) \oplus  \qty(\rep{2}^0,\rep{1}^{-3/2} \oplus \crep{3}^{1/2} ).  \end{tikzcd}
\end{equation}
The $\rep{10}$ of $U(5)$, which survives, decomposes as
$(\mathbf{1}, \mathbf{3}) \oplus (\mathbf{2}, \mathbf{\overline{3}}) \oplus (\mathbf{1}, \mathbf{1}).$
These decompositions can be summarized as shown in Table~\ref{tab:4dred}.
The reader will recognize the fields as the now-familiar algebra
\deq{
\Omega^{0,*}(\C^2)\otimes \C[\psi_1,\psi_2, \psi_3] \otimes \lie{u}(N).
}
We can also view the fields in terms of $\N = 1$ language.
 The fields on the leftmost diagonal correspond to $\Omega^{0,*}(\C^2,  \mathfrak{g})$ which correspond to the $\mathcal{N} = 1$ vector.  The fields on the rightmost diagonal correspond to $\Omega^{2,*}(\C^2,  \mathfrak{g})[1]$, which is the antifield to the $\N = 1$ vector.  Together, these are the fields of holomorphic $BF$ theory, which is the holomorphic twist of pure $\N=1$ gauge theory in four dimensions~\cite{Baulieu:2004pv, Costello:2013zra}. The inner fields correspond to three $\mathcal{N} = 1$ chiral multiplets which transform in the $ \crep{3}$ of the $R$-symmetry group, and their antifields which transform in the $\rep{3}$ of the $R$-symmetry group.
\subsection{Traditional letter counting and interaction spectral sequences}
\label{sec:letter}
We now pause to compare with the traditional letter counting approaches to computing the superconformal index \cite{Romelsberger:2005eg, Kinney:2005ej}.  In the notation of \cite{Kinney:2005ej, Grant:2008sk, Chang:2013fba} the letters that contribute to the superconformal index of  $\N = 4$ super Yang--Mills are
\deq{
\phi^n \equiv \Phi^{4m}, \qquad \psi_n \equiv -i \Psi_{n+}, \qquad \lambda_{\alphadot} = \overline{\Psi}_{\alphadot}^{4}, \qquad f \equiv -i F_{++}.
}
One starts by considering the set of all expressions generated by these letters. Then, to organize the action of $Q$ on the derivatives of fields, the following generating fields are introduced:
\begin{align}
\lambda^m(z) & = \sum_{n = 0}^{\infty} \frac{1}{(n+1)!} (z^{\alphadot} D_{\alphadot})^n (z^{\betadot} \lambda_{\betadot}) \nonumber \\
\phi^m(z) & = \sum_{n = 0}^{\infty} \frac{1}{n!} (z^{\alphadot} D_{\alphadot})^n \phi^m \nonumber \\
\psi_m(z) & = \sum_{n = 0}^{\infty} \frac{1}{n!} (z^{\alphadot} D_{\alphadot})^n \psi_m \\
f(z) & = \sum_{n = 0}^{\infty} \frac{1}{n!} (z^{\alphadot} D_{\alphadot})^n f, \nonumber
\end{align}
where $z^{\alphadot}$ are auxiliary commuting variables.
These generating fields can then be organized in terms of a $(2|3)$ superfield,
\deq{
\Psi(z, \theta) = -i \left[\lambda(z) + 2 \theta_n \phi^n(z) + \epsilon^{mnp} \theta_m \theta_n \psi_p(z) + 4 \theta_1 \theta_2 \theta_3 f(z) \right],
}
The residual action of the supercharge~$Q$ on these generating fields can then be conveniently formulated~\cite{Chang:2013fba} as
\deq[QPsi]{
\{Q, \Psi \} =  \Psi^2.
}
While this Maurer--Cartan-like equation appears somewhat miraculously in the calculations of~\cite{Grant:2008sk, Chang:2013fba}, it follows simply from the form of the supercharge in the dimensionally reduced holomorphic Chern--Simons theory. And it is now perfectly clear what the meaning of the ``residual action'' of the supercharge is: it is just the spectral sequence from the holomorphically twisted free theory to the holomorphically twisted interacting theory~\cite{Gukov:2015gmm}!
In particular, it corresponds to going the ``other way'' around a commuting square, performing the holomorphic twist \emph{before} turning on interactions, rather than after. In this manner, one avoids ever dealing with the untwisted interacting theory. 

In this language, the interpretation of a result like~\eqref{QPsi} is entirely clear. One starts with the BV complex,
\deq{
\left[\Omega^{0,*}(\C^2)\otimes \C[\psi_1,\psi_2, \psi_3] \otimes \lie{u}(N); d\alpha =( d_0 + d_1 )\alpha  = \bar{\partial} \alpha + g \alpha \wedge \alpha \right].
}
One first applies $d_0$; when one does this, $\Omega^{0,*}(\C^2)$ collapses to holomorphic functions $\mathscr{O}(\C^2)$, and the exterior algebra and gauge indices come along for the ride. The result is the next page of the spectral sequence,
\deq{
\mathscr{O}(\C^2)\otimes \C[\psi_1,\psi_2, \psi_3] \otimes \lie{u}(N); d_1^*\alpha = g\alpha \wedge \alpha.
} 
The reader will have no trouble recognizing this as the generating field $\Psi$ and the differential~\eqref{QPsi}, which now has a clear physical meaning. (Again, implicit in all computations like this is the collapse of the spectral sequence at~$E_2$.)

In the previous section, the $c$-ghost plays the role of $\lambda^m(z)$ and the singlet from the gaugino antifield plays the role of $f(z).$    The $F_{+}$-terms appear as singlets of a gaugino antifield.  Their antifields $F_{+}^{\af}$ are represented by $\psi_p$ in letter counting.

\subsection{Dimensional reduction to two dimensions}
\label{sub:2d}
\begin{table}[thp]
\caption{Dimensional reduction of fields to two dimensions}
\[
\begin{tikzcd}[row sep = 1ex, column sep = 1ex]
(1,1) & (1,\mathbf{\overline{4}}) & (1,\mathbf{6}) & (1,\mathbf{4}) & (1,\mathbf{1}) & \\
& (1,1) & (1,\mathbf{\overline{4}}) & (1,\mathbf{6}) & (1,\mathbf{4}) & (1,\mathbf{1})  \\
\hline
\mathbf{1} & \mathbf{\overline{5}} & \mathbf{\overline{10}} & \mathbf{10} &  \mathbf{5} &  \mathbf{1}\\    
\end{tikzcd}
\]
\label{tab:2dred}
\end{table}

By analogy with the discussion in four dimensions, we consider the following diagram of inclusions of groups:
\begin{equation}
\begin{tikzcd}
SO(10)  & SO(2) \times SO(8) \ar[l] \\
U(5) \ar[u] & U(1) \times U(4) \ar[u] \ar[l]. 
\end{tikzcd}
\end{equation}
Note that this is the same diagram of groups as appears in~\cite{Murat}; in that context, $U(4)$ appears due to $R$-symmetry breaking when a $(2,2)$ subalgebra of $\mathscr{N}=(8,8)$ is selected in two dimensions.

The branching rules for the ten-dimensional vector representation are easy to determine:
\begin{equation}
\begin{tikzcd}
\mathbf{10} \ar[r] \ar[d] & \mathbf{2}\oplus\mathbf{8} \ar[d] \\
\mathbf{5}^{1}\oplus\smash{\mathbf{\overline{5}}}^{-1} \ar[r] & (+1,\rep{1}^0)\oplus(0,\rep{4}^1)\oplus (-1,\rep{1}^0) \oplus(0,\crep{4}^{-1})
\end{tikzcd}
\end{equation}
Similarly, the branching rules for the ten-dimensional spinor representation fit into the diagram 
\begin{equation}
\begin{tikzcd}
S_+ \ar[r] \ar[d] & \left(\mathbf{8}_+,{\frac{1}{2}} \right)\oplus \left(\mathbf{8}_-,{-\frac{1}{2}} \right) \ar[d] \\
 \mathbf{5}^{-3/2} \oplus \smash{\crep{10}}^{1/2} \oplus \mathbf{1}^{5/2} \ar[r] & 
 \qty(-\frac{1}{2},\mathbf{4}^{-1} \oplus \crep{4}^{1}) \oplus \qty(+\frac{1}{2}, \mathbf{1}^{-2} \oplus \mathbf{1}^{2} \oplus \mathbf{6}^0)
 \end{tikzcd}
\end{equation}
The fields of twisted super Yang--Mills theory are summarized in Table \ref{tab:2dred}.
From the vector multiplet, we see that there will be a two-dimensional gauge field $A_{\mu}$ which transforms as 
the~$\mathbf{2}$ of $SO(2)$ (i.e., two charged scalars of~$U(1)$).  The remaining eight components are eight scalar fields that transform in the $\mathbf{4}\oplus \mathbf{\overline{4}}$ of the $U(4)$ $R$-symmetry group.  In the dimensionally reduced holomorphic twist, only the scalars in the $\mathbf{\overline{4}}$ will survive. From the fermions, the $\crep{4}$ and the $\rep{6}$ appear in the twisted theory.
Just as in the four-dimensional case, the fields can be neatly summarized with the BV complex
\deq{
\Omega^{0,*}(\C)\otimes \C[\psi_1,\psi_2, \psi_3, \psi_4].
}

\section{$\N=(0,2)$ theories}
\label{sec:0,2}

We now make some remarks about more general $\N=(0,2)$ theories. By way of notation, we use $\N=(0,2)$ superspace following \cite{Witten:1993yc} to construct manifestly supersymmetric Lagrangians.
The bosonic coordinates are $y^{\alpha}$ ($\alpha = 1,2$); fermionic coordinates are $\theta^{+}$ and~$\thetabar^{+}$.
The two supersymmetry generators are
\begin{eqnarray}
\Q_{+} & = & \pdv{}{\theta^{+}} + i \thetabar \left(\pdv{}{y^0} + \pdv{}{y^1} \right) \\
\Qbar_{+} & = & \pdv{}{\thetabar^{+}} - i \theta \left(\pdv{}{y^0} + \pdv{}{y^1} \right).
\end{eqnarray}
The supersymmetry generators commute with the super derivatives
\begin{eqnarray}
D_{+} & = & \pdv{}{\theta^{+}} - i \thetabar \left(\pdv{}{y^0} + \pdv{}{y^1} \right) \\
\overline{D}_{+} & = & \pdv{}{\thetabar^{+}} + i \theta \left(\pdv{}{y^0} + \pdv{}{y^1} \right).
\end{eqnarray}
There are two types of matter multiplets which we now introduce.
The first matter field we consider is the $(0,2)$ chiral multiplet $\Phi$, which obeys
\deq{
\Dbar_{+} \Phi = 0
}
and has components
\deq{
\Phi = \phi + \sqrt{2} \theta^{+} \psi_{+} - i \theta^{+} \thetabar^{+} (D_0 + D_1) \phi.
}
The Fermi multiplet $\Lambda_{-}$ satisfies
\deq{
\Dbar_{+} \Lambda_{-} = \sqrt{2} E,
}
where $E$ is a superfield obeying
\deq{
\Dbar_{+} E = 0.
}
We will only need to consider the case where $E = E(\Phi_i)$ is a holomorphic function of some chiral superfields $\Phi_i$.
In addition to the $E$-interaction, we can introduce a $J$-interaction using the $(0,2)$ analog of the superpotential:
\deq{
L_J = \int d^2y d\theta^{+} \; \Lambda_{-,a} J^{a}(\Phi_i) \vert_{\thetabar^{+} = 0}.
}
For this term to be supersymmetric, the following constraint must be satisfied:
\deq{
\sum_a E_a(\Phi_i) J^a(\Phi_i) = 0.
}
The $E$ and $J$-interactions can be exchanged be replacing the Fermi multiplet $\Lambda_{-}$ with its conjugate $\overline{\Lambda}_{-}$.

\section{Quivers and Calabi--Yau $d$-algebras}
\label{sec:quivCY}

\subsection{Quivers with superpotential}
\label{sec:quivers}

Quivers have been intensely studied in the context of D-branes \cite{Douglas:1996sw}. To recall briefly, a probe D-brane that is pointlike may decay into fractional branes. The gauge theory of open strings between the fractional branes then takes the form of a quiver gauge theory.  The (scalar) moduli space of vacua of the D-brane worldvolume theory encodes the motion of the original D-brane in the Calabi--Yau.  Therefore the moduli space of vacua can be thought of as a non-commutative resolution of the Calabi--Yau singularity \cite{MR2077594}.  For an introduction, explanation, and visualization of the decay of D-branes into fractional branes in the $A$-model, see~\cite{Eager:2016yxd}.

A particular feature of D3-branes at a Calabi--Yau threefold singularity is that all of the relations come from a single function known as the superpotential.  Furthermore, a form of Serre duality results in the quiver with potential having a self-dual resolution as a non-commutative algebra~\cite{Berenstein:2002fi}.  This observation led to the notion of a Calabi--Yau 3-algebra in~\cite{vdbs, Ginzburg:2006fu} and to Calabi--Yau $n$-algebras in~\cite{Ginzburg:2006fu}.  
All exact Calabi--Yau algebras arise from a differential graded (dg)-quiver with superpotential~\cite{MR3338683}.  Early applications of Calabi--Yau $3$-algebras to AdS/CFT appear in~\cite{Eager:2010yu}.

In this section we introduce the structure of exact Calabi--Yau d-algebras following~\cite{Lam:2014} and show that in two dimensions, CY 4-algebras can be used to describe two-dimensional  $\N=(0,2)$ supersymmetric gauge theories probing Calabi--Yau fourfold cones.  Several examples from~\cite{Lam:2014} are given.  More elaborate examples, giving a graphical description of the quivers with potential, appear in~\cite{Craw:2010zi}, using methods adapted from~\cite{Eager:2010ji}.

Several interesting new features appear in two dimensions.  Unlike the situation for toric Calabi--Yau threefolds, there does not always exist an abelian gauge theory description for toric Calabi--Yau fourfolds as the following simple example of 
\cite{SpenkovdBtoric, MR3698338} shows.  A $(\C^{*})^2$ action on $\C^6$ with weights
$(1, 1), (-1, 0), (0, -1), (-3, -3), (3, 0), (0, 3)$ has no non-commutative crepant resolution (NCCR) with only rank one modules.  However, \cite{MR3698338}  shows that there exists a NCCR given by the direct sum of 12 rank-one modules and a single module of rank two.  This corresponds to a quiver gauge theory with gauge group $U(1)^{11} \times U(2).$

\subsection{Calabi--Yau $d$-algebras}
\label{sec:CYd}
A quiver $Q$ can be described as a collection of vertices $Q_0$, arrows $Q_1$ and maps $ h : Q_1 \rightarrow Q_0$ and $t : Q_1 \rightarrow Q_0$ called the ``head'' and ``tail'' of an arrow.  A Calabi--Yau 3-algebra consists of the following data.
\begin{enumerate}
	\item A quiver $Q = \left(Q_0, Q_1, h : Q_1 \rightarrow Q_0, t : Q_1 \rightarrow Q_0\right).$ 
	\item A superpotential $\Phi \in \C Q/[\C Q,  \C Q]$.
\end{enumerate}
where $\C Q$ is the path algebra of the quiver $Q.$

In addition to the edges of a quiver $Q$, a Calabi--Yau 4-algebra has an additional set $R$ of edges.  We will construct a dg algebra where the edges in $Q_1$ will have degree 0 and the edges in $R$ will have degree -1.  A Calabi--Yau 4-algebra consists of the following data.
\begin{enumerate}
	\item A quiver $Q = \left(Q_0, Q_1, h : Q_1 \rightarrow Q_0, t : Q_1 \rightarrow Q_0\right).$ 
	\item Maps $h: R \rightarrow Q_0$ and $t: R \rightarrow Q_0$
	\item A map $A: R \rightarrow \C Q$ such that $A_r := A(r) \in e_{h(r)}  \C Q  e_{t(r)}$.
	\item A symmetric function $q: R \times R \rightarrow \C$ such that
		\begin{enumerate}
		\item $q(r,s) = 0$ unless $r$ and $s$ have the same underlying edge are oppositely oriented.  
			That is to say, $h(r) = t(s)$ and $h(s) = t(r).$ 
		\item $q$ is nondegenerate -- meaning that the matrix $q(r,s)$ is invertible.	
		\item $\sum_{r,s \in Q_1} q(r,s) A_r A_s = 0 \mod \left[\C Q, \C Q \right].$
		\end{enumerate}
\end{enumerate}

\subsection{Calabi--Yau $d$-algebras from quivers}

A {\it quiver with superpotential} of dimension $D$ is a dg-quiver $\widehat{\C Q}$ equipped with a {\it superpotential}
$\Phi \in \widehat{\C Q}_{cyc}$ of degree $3-D$ with additional structure.  The superpotential must satisfy the non-commutative BV-master equation $\left\{ \Phi, \Phi \right\} = 0$.  This structure describes topological B-branes in the B-model \cite{Lazaroiu:2005da}.  
In dimension three, the superpotential is in degree zero.  For CY 3-algebras, the superpotential is the ordinary superpotential of the corresponding four-dimensional quiver gauge theory.  In~\cite{Eager:2012hx}, the differential in the dg algebra associated to a CY 3-algebra was shown to coincide with 
a complex of fields arising from the supersymmetric gauge theory.  An identical phenomenon happens for two-dimensional gauge theories.  In two dimensions, the traditional letter counting reviewed in~\S\ref{sec:letter} becomes somewhat involved.  However, the action of the BV differential $\bvd$ and the scalar supercharge $Q$ on the fields and antifields of the theory in the BV formalism provides a concise way to describe the contribution of fields to the two-dimensional supersymmetric index. This corresponds to the flavored NS elliptic genus.  

Just as in four dimensions, we separate the terms coming from interactions from those appearing in the free theory, so that $d = d_0 + d_1$. The Ginzburg dg algebra associated to the CY $d$-algebra is then isomorphic to the complex $H^{*}(d_0)$, equipped with differential $d_1$.

In four dimensions, the superpotential for a CY $4$-algebra is in degree $-1$.  For CY $4$-algebras, the superpotential is not the superpotential of the 2d theory, but it encodes the $J$ and $E$ interaction terms.  Since the potential is of degree $-1$, each term in the potential is built out of an arbitrary number of chiral fields and a single Fermi field.

From the data of a CY 3-algebra, the Ginzburg dg algebra is constructed as follow.
For every arrow $e \in Q_1$ of degree $0$, we add an arrow $e^{*}$ of degree $-1$ in the opposite orientation.  For every vertex $t \in Q_0$, we add a loop $t^{*}$ of degree $-2$.  The degree-$(-1)$ differential $d$ is given by
\begin{align*}
d(e) & = 0 \\
d(e^{*}) & = \partial_e \Phi \\
d(v^{*}) & = \sum_{e \text{ deg } 0} [e, e^{*}]. 
\end{align*}

From the data of the CY $4$-algebra, we can construct a dg algebra as follows.
For every arrow $e \in Q_1$ of degree $0$, we add an arrow $e^{*}$ of degree -2 in the opposite orientation.  For every vertex $v \in Q_0$, we add a loop $v^{*}$ of degree -3.  The differential $d: \widehat{\C Q} \rightarrow \widehat{\C Q}$ is given by $d = \left\{ \Phi, - \right\}$.  More explicitly, the differential is
$d:  \widehat{\C Q} \rightarrow \widehat{\C Q}$ is
\begin{align*}
d(e) & = 0 \\
d(r) & = A_r \\
d(e^{*}) & = \partial_e \Phi \\
d(v^{*}) & = \sum_{e \text{ deg } 0} [e, e^{*}] + \sum_{r,s \text{deg } 1} q(r,s) r s. \\
\end{align*}

\subsection{Examples}
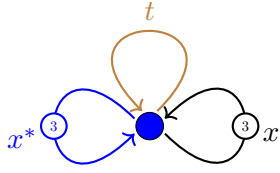
\begin{figure}[ht]
\begin{center}
\begin{tikzpicture}[scale=1] 
\path (0,0) node[draw,shape=circle,fill=blue] (v0) {} node[]{$$}; 
\draw[thick, brown] (.1,.2) edge[above loop] node[above] {$t$} (-.1,.2) ;
\draw[thick, blue] (-.2,.1) edge[leftdown loop] node[midway, draw=blue, shape=circle, fill=white, scale=0.5] {$3$} node[below left, yshift = .7ex, xshift=-.5ex] {$x^{*}$} (-.2,-.1);
\draw[thick] (.2,-.1) edge[right loop] node[midway, draw=black, shape=circle, fill=white, scale=0.5] {$3$} node[below right, yshift = .7ex, xshift=.5ex] {$x$} (.2,.1) ;
\end{tikzpicture}
\caption{dg quiver corresponding to $\C^3$} 
\label{fig:C3}
\end{center}
\end{figure}

The Ginzburg dg quiver corresponding to $\C^3$ is shown in figure \ref{fig:C3}.  This quiver corresponds to $\mathcal{N} = 4$ super Yang-Mills theory in four dimensions.  There are three fields $x$ in degree 0, three $x^{*}$ in degree $-1$ and one field $t$ in degree $-2.$  There is a superpotential
\deq{
\Phi = \epsilon_{ijk} x_i x_j x_k.
}

The three fields $x$ in degree zero correspond to the holomorphic twist of three chiral multiplets of $\mathcal{N} = 4$ super Yang-Mills theory described in section \ref{sub:4d}.  The three
fields $x^{*}$ in degree $-1$ can be viewed either as the twisted anti-fields to the chiral multiplets or as components of fermions in the anti-chiral multiplet.  Finally the field $t$ in degree $-2$ can be viewed as either a particular component $f = F_{++}$ of the gauge field strength or as the dimensional reduction of a particular gaugino antifield.
\begin{figure}[ht]
\begin{center}
\begin{tikzpicture}[scale=1] 
\path (0,0) node[draw,shape=circle,fill=blue] (v0) {} node[]{$$}; 
\draw[thick, brown] (.1,.2) edge[above loop] node[above] {$t$} (-.1,.2) ;
\draw[thick, red] (-.1,-.2) edge[below loop] node[midway, draw=red, shape=circle, fill=white, scale=0.5] {$6$} node[below left, yshift = .7ex, xshift=-.5ex] {$r$} (.1,-.2) ;
\draw[thick, blue] (-.2,.1) edge[leftdown loop] node[midway, draw=blue, shape=circle, fill=white, scale=0.5] {$4$} node[below left, yshift = .7ex, xshift=-.5ex] {$x^{*}$} (-.2,-.1);
\draw[thick] (.2,-.1) edge[right loop] node[midway, draw=black, shape=circle, fill=white, scale=0.5] {$4$} node[below right, yshift = .7ex, xshift=.5ex] {$x$} (.2,.1) ;
\end{tikzpicture}
\caption{dg quiver corresponding to $\C^4$} 
\label{fig:C4}
\end{center}
\end{figure}
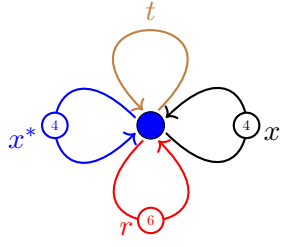

The dg quiver corresponding to $\C^4$ is shown in figure \ref{fig:C4}.  This quiver describes $\mathcal{N} = (8,8)$ super Yang-Mills theory in two dimensions.  There are four fields $x$ in degree 0, six fields $r$ in degree $-1$, four dual fields $x^{*}$ in degree $-2$ and one field $t$ in degree $-3.$
The pairing on degree $-1$ edges is
\deq{
\langle r_{ij}, r_{kl} \rangle = \epsilon_{ijkl},
}
and the superpotential is
\deq{
\Phi = \sum \epsilon_{ijkl}(x_i x_j - x_j x_i) r_{kl}.
}
The four fields $x$ in degree zero correspond to the four $(0,2)$ chiral multiplets described in section \ref{sub:2d}.  Similarly the six fields $r$ in degree $-1$ correspond to Fermi multiplets and their antifields.  The four dual fields $x^{*}$ in degree $-2$ are the antifields to the chiral multiplets.  Finally the field $t$ in degree $-3$ is the antifield to the $(0,2)$ vector multiplet.
\begin{figure}[ht]
\begin{center}
\begin{tikzpicture}[scale=1] 
\path (0,0) node[draw,shape=circle,fill=blue] (v0) {} node[below,yshift = -1ex] {$v_0$}; 
\path (4,0) node[draw,shape=circle,fill=blue] (v1) {} node[below,yshift = -1ex]{$v_1$}; 
\path (8,0) node[draw,shape=circle,fill=blue] (v2) {} node[below,yshift = -1ex]{$v_2$}; 
\triplearrow{arrows={-Implies}}{(.2,-.1) to[bend right=10] node[below]{$a$}  (3.8,-.1)} ; 
\triplearrow{arrows={-Implies}}{(3.8,.1) to[bend right=10] node[above]{$a^{*}$}  (0.2,.1)} ; 
\triplearrow{arrows={-Implies}}{(4.2,-.1) to[bend right=10] node[below]{$b$}  (7.8,-.1)} ; 
\triplearrow{arrows={-Implies}}{(7.8,.1) to[bend right=10] node[above]{$b^{*}$}  (4.2,.1)} ; 
 \path[->, thick, bend right = 30]  (3.8,.3) edge node[above,yshift=-.2ex] {$u$}  (.2,.3)  ;
\path[->, thick, blue, bend left = 50] (.2,.4) edge node[above] {$u^{*}$}  (3.8,.4);
 \path[->, thick, bend right = 30]   (7.8,.3) edge node[above,yshift=-.2ex] {$v$}   (4.2,.3);
\path[->, thick, blue, bend left = 50] (4.2,.4)  edge node[above] {$v^{*}$} (7.8,.4);
 \path[->, thick, bend left = 20]  (7.8,-.6)  edge node[below,yshift=0ex] {$w$} (.2,-.7) ;
\path[->, thick, blue, bend right = 25] (.2,-.9) edge node[below,yshift=.8ex] {$w^{*}$} (7.8,-.9);
\triplearrow{red, arrows={-Implies}}{(.2,-1.2) to[bend right=40] node[below]{$s$}  (7.8,-1.2)} ;
\triplearrow{red, arrows={-Implies}}{ (7.8,.8) to[bend right=40] node[below]{$r$}  (.2,.8)} ;
\draw[thick, red] (v1) edge[above loop] node[midway, draw=red, shape=circle, fill=white, scale=0.5] {$3$} node[above left] {$p$} (v1) ;
\draw[thick, red] (v1) edge[below loop] node[midway, draw=red, shape=circle, fill=white, scale=0.5] {$3$} node[below left, yshift = .7ex, xshift=-.5ex] {$q$} (v1) ;
\path[thick, brown] (v0) edge[left loop] node[left] {$v_0^{*}$} (v0);
\path[thick, brown] (v1) edge[left loop] node[left] {$v_1^{*}$} (v1);
\path[thick, brown] (v2) edge[right loop] node[right] {$v_2^{*}$} (v2);
\end{tikzpicture}
\caption{dg quiver corresponding to $\cO_{\P^2}(-1) \oplus \cO_{\P^2}(-2) \rightarrow \P^2$} 
\label{fig:localp2dg}
\end{center}
\end{figure}
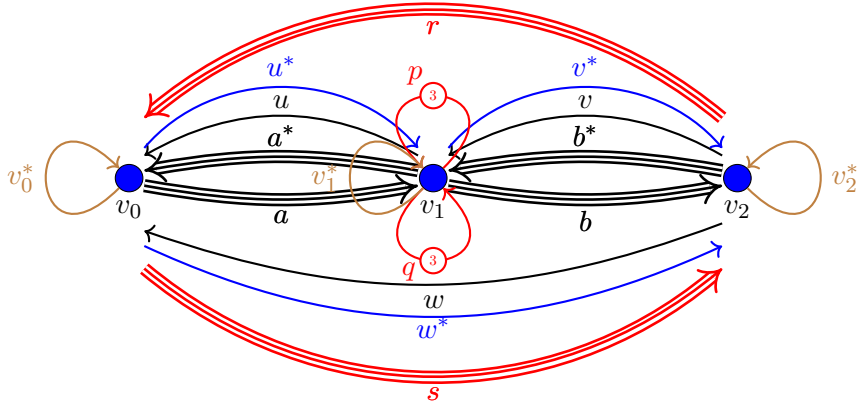

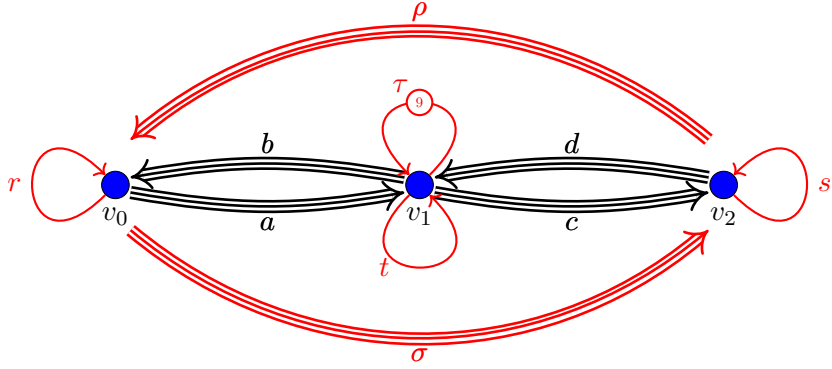
\begin{figure}[ht]
\begin{center}
\begin{tikzpicture}[scale=1] 
\path (0,0) node[draw,shape=circle,fill=blue] (v0) {} node[below,yshift = -1ex] {$v_0$}; 
\path (4,0) node[draw,shape=circle,fill=blue] (v1) {} node[below,yshift = -1ex]{$v_1$}; 
\path (8,0) node[draw,shape=circle,fill=blue] (v2) {} node[below,yshift = -1ex]{$v_2$}; 
\triplearrow{arrows={-Implies}}{(.2,-.1) to[bend right=10] node[below]{$a$}  (3.8,-.1)} ; 
\triplearrow{arrows={-Implies}}{(3.8,.1) to[bend right=10] node[above]{$b$}   (.2,.1)} ; 
\triplearrow{arrows={-Implies}}{(4.2,-.1) to[bend right=10] node[below]{$c$}  (7.8,-.1)} ; 
\triplearrow{arrows={-Implies}}{(7.8,.1) to[bend right=10] node[above]{$d$}   (4.2,.1)} ; 
\triplearrow{red, arrows={-Implies}}{(.2,-0.6) to[bend right=40] node[below]{$\sigma$}  (7.8,-0.6)} ;
\triplearrow{red, arrows={-Implies}}{ (7.8,.6) to[bend right=40] node[above]{$\rho$}  (.2,.6)} ;
\draw[thick, red] (v1) edge[above loop] node[midway, draw=red, shape=circle, fill=white, scale=0.5] {$9$} node[above left] {$\tau$} (v1) ;
\draw[thick, red] (v1) edge[below loop] node[left, xshift=-1.5ex] {$t$} (v1) ;
\path[thick, red] (v0) edge[left loop] node[left] {$r$} (v0);
\path[thick, red] (v2) edge[right loop] node[right] {$s$} (v2);
\end{tikzpicture}
\caption{Underlying quiver corresponding to $T_{\P^2}^{\vee} \rightarrow \P^2$} 
\label{fig:localp2b}
\end{center}
\end{figure}

Several examples of quivers corresponding to $\text{CY}^4$ varieties are given in \cite{Lam:2014}.  We now recall two of them to give a flavor of what these theories look like.  The dg quiver for $\cO_{\P^2}(-1) \oplus \cO_{\P^2}(-2) \rightarrow \P^2$ \cite{Lam:2014} is shown in figure \ref{fig:localp2dg}.  The superpotential is
\deq{
\Phi = (a_{i + 2} w b_{i + 1} - a_{i + 1} w b_{i + 2})p_i + (a_i u - v b_i) q_i + (b_{i + 2} a_{i +1} - b_{i + 1} a_{i+2}) r_i + (w b_i v - u a_i w) s_i.
}
The quiver for $T_{\P^2}^{\vee} \rightarrow \P^2$ \cite{Lam:2014} is quite complicated, so we only display the fields in degree $0$ and $-1$ in figure \ref{fig:localp2b}.
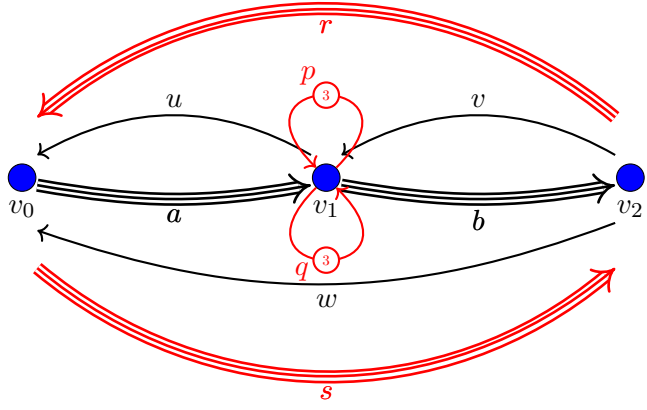
\begin{figure}[ht]
\begin{center}
\begin{tikzpicture}[scale=1] 
\path (0,0) node[draw,shape=circle,fill=blue] (v0) {} node[below,yshift = -1ex] {$v_0$}; 
\path (4,0) node[draw,shape=circle,fill=blue] (v1) {} node[below,yshift = -1ex]{$v_1$}; 
\path (8,0) node[draw,shape=circle,fill=blue] (v2) {} node[below,yshift = -1ex]{$v_2$}; 
\triplearrow{arrows={-Implies}}{(.2,-.1) to[bend right=10] node[below]{$a$}  (3.8,-.1)} ; 
\triplearrow{arrows={-Implies}}{(4.2,-.1) to[bend right=10] node[below]{$b$}  (7.8,-.1)} ; 
 \path[->, thick, bend right = 30]  (3.8,.3) edge node[above,yshift=-.2ex] {$u$}  (.2,.3)  ;
 \path[->, thick, bend right = 30]   (7.8,.3) edge node[above,yshift=-.2ex] {$v$}   (4.2,.3);
 \path[->, thick, bend left = 20]  (7.8,-.6)  edge node[below,yshift=0ex] {$w$} (.2,-.7) ;
\triplearrow{red, arrows={-Implies}}{(.2,-1.2) to[bend right=40] node[below]{$s$}  (7.8,-1.2)} ;
\triplearrow{red, arrows={-Implies}}{ (7.8,.8) to[bend right=40] node[below]{$r$}  (.2,.8)} ;
\draw[thick, red] (v1) edge[above loop] node[midway, draw=red, shape=circle, fill=white, scale=0.5] {$3$} node[above left] {$p$} (v1) ;
\draw[thick, red] (v1) edge[below loop] node[midway, draw=red, shape=circle, fill=white, scale=0.5] {$3$} node[below left, yshift = .7ex, xshift=-.5ex] {$q$} (v1) ;
\end{tikzpicture}
\caption{Underlying quiver corresponding to $\cO_{\P^2}(-1) \oplus \cO_{\P^2}(-2) \rightarrow \P^2,$ from \cite{Lam:2014}.} 
\label{fig:localp2a}
\end{center}
\end{figure}

The superpotential is 
\begin{align*}
\Phi = \sum_{i = 0}^{2}  &  (a_i b_i + d_i c_i)t - 2(b_i a_i)r - 2(c_i d_i) s + 2(c_{i + 2} a_{i + 1} - c_{i+1}a_{i+2}) \sigma_i + 2 (b_{i+2} d_{i+1} - b_{i+1} d_{i+2}) \rho_i \\
& + 2(a_i b_{i+1} - d_{i+1}c_i)\tau_{i,i+1} + 2(a_i b_{i-1} - d_{i-1} c_{i} \tau_{i, i-1} + (a_i b_i - d_i c_d)(\tau_{ii} - \tau_{i+1, i+1}- \tau_{i-1, t-1}).  \\
\end{align*}

\section{Equivariant DMVV formulae}
\label{sec:DMVV}

While the cohomology problem for finite-dimensional gauge groups is difficult, there are drastic simplifications in the large-$N$ limit.  
The large-$N$ single-trace index can be computed in terms of cyclic homology of the Ginzubrg dg algebra \cite{Eager:2012hx}.  Since there are no trace relations in the large-$N$ limit, the full superconformal index can be determined by a Fock-space construction out of the single trace index.  The result is that
\deq{
\mathscr{I}_{s.t} =\chi(HC^{\bullet}).
}
The result relies on a detailed saddle-point analysis of a matrix model \cite{Aharony:2003sx}.  Assuming that a similar computation will work in two dimensions, we have the conjectured form of the large-$N$ flavored elliptic genus
\deq{
\mathscr{I}_{s.t} = \chi(HC^{\bullet}).
}
We defer the details of this computation to future work.  A limiting form of the flavored elliptic genus is the Hirzebruch $\chi_y$ genus.  

In this section we will show for $Y = \C^k$ that Hirzebruch $\chi_y$ genus of the infinite symmetric product of $Y$ can be expressed in terms of the cyclic homology of $Y$ or its corresponding quiver.
We use an equivariant version of the results of DMVV \cite{Dijkgraaf:1996xw} formulated in \cite{MR2213685} which relates the elliptic genus of an infinite symmetric product of $Y$ to the elliptic genus of $Y$ itself.  See \cite{Dijkgraaf:1998zd} for an elegant review.  An interesting related result is the description of the $S^1$-equivariant cohomology of the loop space of $Y$ in terms of the cyclic homology of $Y$ \cite{MR870737}.

\subsection{Equivariant $\chi_y$ genus}
Let $X$ be a complex $d$-dimensional manifold.  For any holomorphic vector bundle $V$ over $X$, its Euler character is defined to be
\deq{
\chi(X, V) = \sum_{q \ge 0} (-1)^q \dim H^q(M, V).
}
We define the formal sum
\deq{
\Lambda_q = \bigoplus_{k \ge 0} q^k \Lambda^k V,
}
where $\Lambda^k$ denotes the $k$-th exterior power.
Recall that the Hirzebruch $\chi_y$ genus of $X$ is given by
\begin{align}
\chi_y(X) & = \sum_{p = 0}^{d} (-y)^p \sum_{q = 0}^{d} (-1)^q \dim H^q(X, \Lambda^p T^{*} X) \nonumber \\
 & = \sum_{p \ge 0} y^p \chi(X, \Lambda^p T^{*} X).
\end{align}
Let $x_1, \dots, x_d$ denote the formal Chern roots of the holomorphic tangent bundle $TX.$
Then the $\chi_y$ genus can be evaluated using the Hirzebruch Riemann-Roch theorem, 
\begin{align}
\chi_{-y}(X) & = \int_{X} \ch \Lambda_{-y}(T^{*} X) \text{Td}(X) \nonumber \\
& = \int_{X}\, \prod_{j =1}^{d} (1 - y e^{-x_j}) \frac{x_j}{1 - e^{-x_j}},
\end{align}
where $\text{Td}(X)$ is the Todd class of $X$.
For non-compact $X$, the $\chi_y$ genus is often ill defined.  In this case, we consider the equivariant version following \cite{MR1953295, MR2213685}.  Suppose now that $X$ admits a $G$-action and the holomorphic vector bundle $V$ admits a $G$-action which is compatible with the $G$-action on $X.$  Then, for $g \in G$, the Lefschetz number is defined to be
\deq{
L(X, V)(g) = \sum_{i = 0}^{\dim M} (-1)^{i} \Tr(g \vert_{H^i(X, \cO(V))}).
}
The Lefschetz number can be computed by the holomorphic Lefschetz formula
\deq{
L(X, V)(g) = \int_{X^{g}} \frac{\text{ch}_{g} V}{\text{ch}_{g} \Lambda_{-1}(N_{M^g/M})},
}
where $M^g$ denotes the fixed point set of $g$ and $N_{M^g/M}$ is the normal bundle of $M^g$ in $M.$
The equivariant orbifold elliptic genus is defined to be
\deq{
\chi_y(M,G,q, y)(t) = \sum_{[g] \in G_{*}} \frac{1}{|Z(g)|} \sum_{h \in Z(g)} L(M^g, E(M; q,y)^g)(h,t),
}
where $E(M; q,y)^g$ is a virtual bundle defined in \cite{MR1953295, MR2213685}, the sum is over conjugacy classes $G_{*}$ of $G,$ and $Z(g)$ denotes the centralizer of $g$ in $G.$
The equivariant orbifold Hirzebruch $\chi_y$ genus is then defined to be the limit of the elliptic genus where $q$ goes to zero.  These formulas are simply mathematical incarnations of the usual calculations in orbifold conformal field theory.
\subsection{The large-$N$ limit of the equivariant $\chi_y$ genus}
Let $t_1, t_2, \ldots, t_d$ be equivariant parameters for the coordinates of $\C^d.$  The equivariant $\chi_y$ genus of $\C^d$ is
\deq{
\chi_y(\C^d)(t_1, t_2, \dots, t_d) = \prod_{j = 1}^{d} \frac{(1 - y t_j)}{(1-t_j)}.
}
For $\C^2$ the equivariant $\chi_y$ genus specializes to
\deq{
\chi_y(\C^2) = \frac{(1 - y t_1)(1- y t_2)}{(1-t_1)(1-t_2)}.
}
This can be seen by direct computation.
\deq{
H^q(\C^2, \Lambda^p T^{*} \C^2) = 
\begin{cases}
\C[z_1,z_2] & (p,q) = (0,0), \\
\C[z_1,z_2] dz_1 \oplus \C[z_1,z_2] dz_2  & (p,q) = (1,0), \\
\C[z_1,z_2] dz_1 \wedge dz_2  & (p,q) = (2,0), \\
0 & \text{otherwise.}
\end{cases}
}
The contribution from $(p,q) = (0,0)$ is the equivariant Hilbert series
\deq{
\chi_0(t_1,t_2) = \frac{1}{(1 - t_1)(1 - t_2)}.
}
The contributions from $(p,q) = (1,0)$ and $(2,0)$
are
\deq{
\chi_0(t_1,t_2)(-yt_1 - y t_2) \text{ and } \chi_0(t_1,t_2)(y^2 t_1 t_2),
}
respectively.  Summing all the contributions results in
\begin{align}
\chi_y(\C^2) & =  \chi_0(t_1,t_2)(1 - y t_1 - y t_2 + y^2 t_1 t_2) \nonumber\\
& = \frac{(1 - y t_1)(1- y t_2)}{(1-t_1)(1-t_2)}. 
\end{align}
Using the $\chi_y$ genus of $\C^d$ we can compute the equivariant $\chi_{y}$ genus of the large-$N$ symmetric product of $\C^d$ by first constructing a generating series $\sum_{n \ge 0} Q^n \chi_y((\C^d)^n, S_n)(t_1,t_2, \dots, t_d)$ of all the symmetric products and then extracting the $n \rightarrow \infty$ limit.  For $\C^2$ the generating series is given by~\cite{MR2213685}
\deq{
 \sum_{n \ge 0} Q^n \chi_y((\C^2)^n, S_n) 
 =  \prod_{l \ge 1} \prod_{m_1,m_2 \ge0} \frac{(1 - y^l Q^l t_1^{m_1 + 1} t_2^{m_2})(1 - y^l Q^l t_1^{m_1} t_2^{m_2 + 1})}{(1 - y^{l - 1} Q^l  t_1^{m_1} t_2^{m_2})(1 - y^{l + 1} Q^l  t_1^{m_1+1} t_2^{m_2 + 1})} .
}
We can extract the $n \rightarrow \infty$ limit by multiplying the generating function by $(1 - Q)$ and taking the limit $Q \rightarrow 1$ \cite{Kinney:2005ej}.  It is easy to see that all terms with powers of $y$ greater than one vanish.
The plethystic log is given by
\begin{align}
t_1^{m_1} t_2^{m_2} & \qquad   (m_1, m_2) \neq (0,0), l = 0 \nonumber \\
- t_1^{m_1} t_2^{m_2} y & \qquad m_1 \ge 1, m_2 \ge 1, l =1 \nonumber \\
y & \qquad (m_1, m_2) = (0,0), l =1 \\
& \qquad 0 \text{ otherwise.} \nonumber
\end{align}
In general for $\C^d$ all powers of $y$ greater than $d-1$ vanish.
Similarly for $\C^4$
\begin{multline}
\sum_{n \ge 0} Q^n \chi_y((\C^4)^n, S_n) 
=  
\\
\prod_{l \ge 1} \prod_{m_1,m_2 m_3, m_4 \ge0} \frac{(1 - y^l Q^l t^{m + e_i}) (1 - y^{l+2} Q^l t^{m + e_i + e_j + e_k})}{(1 - y^{l - 1} Q^l  )(1 - y^{l + 1} Q^l  t^{m + e_i + e_j} ) 
(1 - y^{l + 3} Q^l  t^{m + e_i + e_j + e_k + e_l} )} ,
\end{multline}
where $t^m = t_1^{m_1} t_2^{m_2} t_3^{m_3} t_4^{m_4}$ and $e_i$ are unit vectors in the space of exponents of $m$.  The large-$N$ orbifold $\chi_y$ genus of $\C^4$ is similar to the $\C^2$ case.  All powers of $y$ greater than three vanish.  However the combinatorics is slightly more involved, so it is worthwhile to pause to introduce some notation following \cite{Eager:2013mua}.

Each holomorphic function $f$ on $\C^4$ has integer charges $m=(m_1,m_2,m_3,m_4)$ under the four isometries, and contributes  
$t^m=t_1{}^{m_1}t_2{}^{m_2} t_3{}^{m_3}t_4{}^{m_4}$  to the equivariant Hilbert series. The charges form a cone $M \subset \bZ^4$ with $m_1, m_2, m_3, m_4 \ge 0.$  Since we are considering $\C^4$, the cone is the standard orthant.
The equivariant Hilbert series is 
 \begin{equation}
\left. \Tr t^m \right|_{H^0(\C^4,\cO_X)}= \sum_{m \in M} t^m .
\end{equation} 
The contribution of order $y^k$ is 
\deq{
(-y)^k \sum_{m \in M} n^{k}_{m} t^m.
}
where
\begin{equation}
n_m^k =\begin{cases}
0 & \text{if $m$ is on a vertex  of $M$},\\
\binom{r-1}{k} & \text{if $m$ is on a $r$-dimensional facet of $M$}.\\
\end{cases}
\end{equation}
These contributions are identified with characters of reduced differential forms on $\C^4$ in \cite{Eager:2012hx, Eager:2013mua}.
The dimensions (graded by total degree) of the cyclic homology groups of $\C^4$ are shown in Table \ref{tab:cyc}.  This provides a geometric interpretation of the $Q + \bvd$ cohomology groups that we have previously computed and hence a geometric description of holomorphic observables.
\begin{table}[th]
\begin{center}
\begin{tabular}{|p{5ex}|p{5ex}|p{5ex}|p{5ex}|p{5ex}|p{5ex}|p{5ex}|p{5ex}|p{5ex}|}
\hline
 & $1$ & $t$ & $t^2$ & $t^3$ & $t^3$ & $t^4$ & $t^5$ & $\dots$ \\
\hline
$HC_{0}$ & 0 & 4 & 10 & 20 & 35 & 56 & 84 & $\dots$ \\
$HC_{1}$ & 0 & 0 & 6 & 20 & 45 & 84 & 140 & $\dots$ \\
$HC_{2}$ & 0 & 0 & 0 & 4 & 15 & 36 & 70 & $\dots$ \\
$HC_{3}$ & 0 & 0 & 0 & 0 & 1 & 4 & 10 & $\dots$ \\
\hline
$\chi(t)$ & 0 & 4 & 4 & 4 & 4 & 4 & 4 & $\dots$ \\
\hline
\end{tabular}
\caption{Cyclic homology group dimensions for $\C^4.$}
\label{tab:cyc}
\end{center}
\label{default}
\end{table}%

\bibliographystyle{ytphys}
\bibliography{CY4}
\end{document}